\documentclass{aa}  

\usepackage[colorlinks=true,citecolor=blue,linkcolor=blue,urlcolor=blue]{hyperref}
\usepackage{graphicx}
\usepackage{txfonts}
\usepackage{lipsum}
\usepackage{xcolor}
\usepackage{subcaption}         
\usepackage{lscape}             
\usepackage{placeins}           

\usepackage{natbib,twoopt}
\bibpunct{(}{)}{;}{a}{}{,} 

\begin{document}

   \title{PAHSPECS: JWST/MIRI Spectroscopy of PAHs at Cosmic Noon}

   \subtitle{}

%

   \author{Irene Shivaei\inst{1}\fnmsep\thanks{Corresponding author: ishivaei@cab.inta-csic.es}
        \and 
    {Tanio D{\'i}az Santos\inst{2}}
        \and
    {Leindert A. Boogaard}\inst{3}
        \and
    {Karin Sandstrom}\inst{4}
        \and
    {Cristina Lofaro}\inst{2}
        \and
    {Fergus R. Donnan}\inst{4}
        \and
    {Pablo G. P{\'e}rez Gonz{\'a}lez}\inst{1}
        \and
    {Roman Fern{\'a}ndez Ar{\'a}nda}\inst{1}
        \and
    Javier {\'A}lvarez M{\'a}rquez\inst{1}
    \and
    Manuel Aravena\inst{5,6} 
        \and
    Roberto Decarli\inst{7}
        \and
    Fr{\'e}d{\'e}ric Galliano\inst{8}
        \and
    Hanae Inami\inst{9}
        \and
    G{\"o}ran {\"O}stlin\inst{10}
        \and
    Gerg\H{o} Popping\inst{11}
        \and
    Fabian Walter\inst{12}
        \and
    Paul van der Werf\inst{3}
    }
   \institute{
   Centro de Astrobiolog\'ia (CAB), CSIC-INTA, Carretera de Ajalvir km 4, Torrej\'on de Ardoz 28850, Madrid, Spain
   \and
   Institute of Astrophysics, Foundation for Research and Technology--Hellas (FORTH), Heraklion, GR-70013, Greece
   \and
   Leiden Observatory, Leiden University, PO Box 9513, 2300 RA Leiden, The Netherlands
   \and
   Department of Astronomy \& Astrophysics, University of California San Diego, 9500 Gilman Drive, San Diego, CA 92093, USA
   \and
   Instituto de Estudios Astrof\'{\i}cos, Facultad de Ingenier\'{\i}a y Ciencias, Universidad Diego Portales, Av. Ej\'ercito 441, Santiago, Chile
   \and
   Millenium Nucleus for Galaxies (MINGAL)
   \and
   INAF Osservatorio di Astrofisica e Scienza dello Spazio di Bologna, via Gobetti 93/3, Bologna, 40129, Italy
   \and
   Université Paris-Saclay, Université Paris Cité, CEA, CNRS, AIM, 91191, Gif-sur-Yvette, France
   \and
   Hiroshima Astrophysical Science Center, Hiroshima University, 1-3-1 Kagamiyama, Higashi-Hiroshima, Hiroshima 739-8526, Japan
   \and
   Department of Astronomy, Oscar Klein Centre, AlbaNova University Center, Stockholm University, 10691 Stockholm, Sweden
   \and
   European Southern Observatory, Karl-Schwarzschild-Str. 2, D-85748, Garching, Germany    
   \and
   Max Planck Institut f\"ur Astronomie, K\"onigstuhl 17, D-69117 Heidelberg, Germany
   }

   \date{Submitted to A\&A}


 \abstract{
The physical conditions of the interstellar medium (ISM) are thought to differ greatly in high-redshift environments compared to those in the local Universe.
Polycyclic aromatic hydrocarbons (PAHs) are one of the most important constituents of the ISM, usually mediating the heating and cooling of gas in photo-dissociation regions. However, PAHs are currently essentially unexplored in main-sequence (MS) galaxies at high redshifts, and specifically at $z$\,$\sim$\,1--2 (‘cosmic noon’), preventing us from accessing the ISM's thermal and chemical balance. Here we present the PAH Spectroscopic Survey (PAHSPECS) JWST program, the first MIRI/MRS survey of PAH features at 3.3, 6.2, 7.7, 8.6 and 11.3\,$\mu$m in MS galaxies at $z$\,$\sim$\,1. The five targets are from the deep ALMA CO and continuum flux-limited sample in the HUDF.
Here, we provide an overview of the project, its motivations, and a brief summary of the first results. We detect the 6.2, 7.7 and 11.3\,$\mu$m PAHs in all galaxies, and the 3.3\,$\mu$m feature in only two. We find that spectral energy distribution (SED) fitting to multi-wavelength broad-band mid-IR photometry can recover the true 6.2 and 7.7\,$\mu$m PAH luminosities derived via spectral decomposition of the MRS spectra to within $\sim$\,25\% accuracy. However, the 3.3 and 11.3\,$\mu$m PAH features are both overestimated with respect to the true values by $\sim$\,50\% (and in some cases up to factors of 3--5). Spatially resolved maps of ASPECS-6, the most extended source in the sample, show significant spatial variations of the 7.7\,$\mu$m PAH contribution to the MIRI F1500W flux, suggesting that while photometric measurements can infer the PAH luminosity accurately, IFU spectroscopy is necessary to study variations in PAH ratios and therefore characterize the dust properties of galaxies. The fractional contribution of PAHs to the total dust mass ($q_{\rm PAH}$) derived from photometric SED modeling of the PAHSPECS sources is roughly in agreement with the $L_{\rm PAH}/L_{\rm IR}$ ratio measured from spectroscopy, but the correlation is not as tight as predicted by the models. On average, PAHSPECS galaxies show an excess in their 6.2\,$\mu$m equivalent width with respect to nearby dusty luminous systems. On the other hand, on average, they have lower 3.3\,$\mu$m PAH equivalent widths than local systems, but they show values similar to other higher mass, starbursting systems at similar redshifts ($z$\,$\sim$\,1--3). This evidence points to differential changes in the size distributions of the neutral and ionized PAH populations in cosmic noon galaxies. 
}

   \maketitle
   \nolinenumbers

\section{Introduction}

A large fraction of the stars in today’s Universe were formed at $z$\,$\sim$\,1--2, a period now coined as cosmic noon. To understand the rapid assembly of stellar mass during this period, it is crucial to obtain a comprehensive picture of the conditions within the interstellar medium (ISM), the constituent of galaxies out of which stars form and into which they inject energy and enriched matter. The conditions within the multi-phase ISM have been probed in great detail in the local Universe via observations of various dust components as well as molecular, atomic and ionized gas \citep{Kennicutt2012, Galliano2018}, but this multi-wavelength view is still missing at $z$\,$\sim$\,1--2. Studies of ionized gas indicate that galaxies at cosmic noon have lower gas metallicities, higher electron densities and harder ionizing radiation fields than the galaxies of the same stellar mass at $z$\,$\sim$\,0 \citep[e.g.,][]{Steidel2014, Kaasinen2017, Strom2017, Sanders2021}, which also result in different dust properties \citep[e.g.,][]{Santini2014, Shivaei2017, Shivaei2022b, Aravena2020, Jolly2025}. Studies of cold gas tracers show that typical star-forming galaxies at cosmic noon also have significantly larger molecular gas reservoirs than their local counterparts \citep[e.g.,][]{Scoville2017, Tacconi2018, Boogaard2023}. However, observations of the atomic neutral gas, which connects the cold molecular and hot ionized phases at cosmic noon are still sparse. This intermediate phase includes the photodissociation regions (PDRs), where far-UV radiation interacts with neutral gas and dust, providing an important link between the physical conditions of the gas and the properties of the interstellar dust.

The cycling of matter in the ISM is a complex, multi-phase process, in which dust plays a crucial role. As key components of the dust population, Polycyclic Aromatic Hydrocarbons (PAHs) account for up to $\sim$\,5\% of the interstellar carbon \citep{Tielens2005}. PAHs are transiently heated by UV photons and reradiate in the mid-infrared (mid-IR), producing distinct emission features. Not only are PAHs reprocessing an important fraction of starlight, with their emission accounting for up to 20\% of the galaxies’ total IR luminosity \citep{Smith2007}, they have been proposed as the main source of heating of the neutral and molecular hydrogen, via the photoelectric effect \citep{Helou2001}. In addition, PAHs act as catalysts for the formation of the H$_2$ molecule, the dominant fuel for star formation. The effectiveness of PAHs in regulating the thermal and chemical balance of the ISM depends on their properties, particularly their charge, size distribution, and abundance relative to larger dust grains \citep{Tielens2008}.

In the local Universe, decades of work have revealed the spatial distribution and relative strengths of different infrared PAH bands from 3 to 17$\mu$m, reflecting a variety of conditions of the ISM across kpc galactic scales \citep[e.g.,][]{Madden2006, Smith2007, DS2008, Galliano2008, DS2010a, DS2010b, DS2011, PS2010a, Chastenet2023, Chown2024, Whitcomb2024, Rigopoulou2024, Baron2025, Lofaro2026a}. In particular, a key result has emerged in the local Universe, in which the main sequence galaxy population tend to show a radial trend where the grain size distribution shifts towards smaller PAHs at large galactocentric radii, consistent with their observed negative metallicity gradients \citep{Whitcomb2024}. 
This suggests PAH growth may be inhibited due to a reduced fraction of carbon available in the gas phase, which impedes the formation of large PAH molecules \citep{Sandstrom2012, Whitcomb2024}. Indeed, a recent, spatially resolved study of the nearby blue compact dwarf II\,Zw\,40 suggests that PAH size distribution is likely shaped by two main mechanisms in low-metallicity environments, namely photo-destruction and inhibited grain growth \citep{Hunt2010,Lai2025}. Changes in PAH ionization state have also been observed in nearby galaxies, especially when strong ionizing sources are present as in the case of active galactic nuclei (AGN). In these extreme environments, contrary to intuition, PAH emission seems to be dominated by the neutral population \citep[e.g.,][]{GB2024}.

Such thorough multi-wavelength studies are still missing beyond the local Universe, where ISM conditions are very different, suggesting different dust properties.
While great progress has been recently made with Atacama Large Millimeter/submillimeter Array (ALMA) to provide an unbiased census of molecular gas and cold dust at high redshift, the rest-frame mid-IR emission of cosmic noon galaxies has been largely unexplored due to the sensitivity, resolution, and spectroscopic coverage limitations of previous telescope facilities. Prior to the James Webb Space Telescope (JWST), PAH emission had only been detected spectroscopically at modest signal-to-noise with Spitzer in a few tens of ultra-luminous infrared galaxies (ULIRGs) and sub-millimeter galaxies (SMGs) beyond $z$\,$\gtrsim$\,1 \citep{Sajina2007, Pope2008, Sajina2009, Siana2009, MD2009, Riechers2014b}. 
JWST can now observe $z>1$ galaxies at much higher fidelity.  Initial studies have explored the 3.3\,$\mu$m PAH feature in a sample of $z$\,$\sim$\,1--3, massive, starbursting galaxies \citep{McKinney2026} and even detected the PAHs 3.3\,$\mu$m at higher redshift in a lensed system \citep{Spilker2023}, while \cite{Wang2026} studied resolved maps of main PAH features at $z=0.9$. Here, we use all wavelength bands of JWST/mid-infrared instrument (MIRI) medium-resolution spectrometer (MRS) to observe the rest-frame near and mid-IR PAH emission of an unbiased sample of normal, star-forming galaxies at $z$\,$\sim$\,1, the peak of cosmic star formation density.

This paper introduces the PAH Spectroscopic Survey (PAHSPECS), a survey of five galaxies selected from the deep, blind ALMA Spectroscopic Survey \citep[ASPECS;][]{Walter2016, Decarli2019, Aravena2020} in the Hubble Ultra Deep Field (HUDF). PAHPECS contains all sources in ASPECS located at $z\sim 1.1$, a redshift that allows to target the 3.3, 6.2, 7.7, 8.6 and 11.3\,$\mu$m PAH features within the MIRI MRS spectral coverage. 
The unique aspect of this program is covering all the mentioned PAH features in a sample of main-sequence galaxies that is complete in CO/dust-continuum flux (i.e., not selected based on their star forming properties). In other words, the PAHSPECS sample, while small, is nevertheless unbiased and fully representative of CO (i.e., gas-) and dust-selected galaxies at $z\sim 1.1$, making it ideal to study the content and properties of PAHs in galaxies at this epoch.

This paper is organized as follows: in Section~\ref{sec:SampleandData}, we describe the sample selection and the newly obtained observations; in Section~\ref{sec:SEDfitting}, we detail the spectral energy distribution modeling performed using broad-band photometry and how it compares with the galaxies' integrated MRS spectra; in Section~\ref{sec:Results}, we discuss the main global results from the program; in Section~\ref{sec:ScienceGoals}, we describe the motivation behind the main works to be released as part of the PAHSPECS collaboration; finally, in Section~\ref{sec:Summary}, we summarize the achieved milestones and we provide prospects for future scientific directions.

The calculations in this paper assume a flat-$\Lambda$CDM cosmology with values $\Omega_{\rm M}$\,=\,0.3 and $H_0$\,=\,70\,km\,s$^{-1}$. At $z$\,=\,1, an angular diameter distance of 1$''$ is equivalent to $\simeq$\,8\,kpc.

\section{Sample and Data}
\label{sec:SampleandData}

\begin{table*}[]
\centering
\footnotesize
\caption{PAHSPECS Sample Properties}
\label{tab:pahspecs_sample_props}
\begin{tabular}{llcccccccc}
\hline\hline
Name  & 3mm ID / 1mm ID  & RA & Dec & $\rm z_{spec}$ & $\log L_{\mathrm{IR}}$ & $\log M_{*}$  & SFR$_{\rm SED}$  & $\tau_{V}$   & $r_{e}^{15\,\mu\mathrm{m}}$, $r_{e}^{1.6\,\mu\mathrm{m}}$\\
      & & (J2000) & (J2000) &  & $[L_{\odot}]$ & $[M_{\odot}]$ & $[M_{\odot}\,\mathrm{yr}^{-1}]$ & [mag] & [arcsec] \\
\hline
ASPECS-6  & 3mm.06 \,/\, 1mm.C16  & 03:32:39.87  & $-27$:47:15.2  & 1.0952  & $11.43^{+0.03}_{-0.02}$  & $10.82^{+0.05}_{-0.05}$ & $28.9^{+1.9}_{-2.4}$  & $0.59^{+0.05}_{-0.05}$ & 0.63, 0.66 \\
ASPECS-11 & 3mm.11 \,/\, --       & 03:32:39.81  & $-27$:46:53.5  & 1.0964  & $10.80^{+0.09}_{-0.06}$  & $10.44^{+0.04}_{-0.04}$ & $6.8^{+0.9}_{-0.7}$   & $0.49^{+0.09}_{-0.07}$ & 0.41, 0.33 \\
ASPECS-14 & 3mm.14 \,/\, 1mm.C25  & 03:32:34.85  & $-27$:46:40.6  & 1.0982  & $11.44^{+0.03}_{-0.03}$  & $10.65^{+0.06}_{-0.06}$ & $22.4^{+2.4}_{-1.9}$  & $1.25^{+0.06}_{-0.06}$ & 0.40, 0.22 \\
ASPECS-15$^{\dagger}$ & 3mm.15 \,/\, 1mm.C12  & 03:32:36.48  & $-27$:46:31.8  & 1.0931  & $11.58^{+0.02}_{-0.02}$  & $10.26^{+0.07}_{-0.10}$  & $35.5^{+3.3}_{-2.2}$  & $1.50^{+0.08}_{-0.06}$ & 0.63, 0.73 \\
ASPECS-C20  & \phantom{3mm..}-- \,/\, 1mm.C20 & 03:32:35.77  & $-27$:46:27.6  & 1.0963  & $11.18^{+0.05}_{-0.04}$  & $10.98^{+0.06}_{-0.05}$  & $10.5^{+1.7}_{-2.4}$  & $1.10^{+0.10}_{-0.08}$ & 0.47, 0.46 \\
\hline
\end{tabular}
\begin{minipage}{0.95\textwidth}
\vspace{0.25cm}
\footnotesize
\textbf{Notes.} Columns:
(1) Source name. (2) ASPECS IDs in 3mm CO and 1mm continuum as defined in the catalogs by \cite{Boogaard2020}, \cite{Aravena2020} and \cite{GL2020}. Refer to these papers for molecular gas mass properties. Columns 2-5 are compiled from these papers. (3)-(4) Coordinates. (5) Redshifts from ALMA CO \citep{GL2020} and MUSE spectroscopy \citep{Boogaard2020}. (6) Infrared Luminosity (L$_{IR}$). (7) Surviving s tellar mass (M$_{\star}$). (8) SED inferred SFR averaged over 100\,Myr. (9) SED inferred V-band optical depth. (10) Effective (half-light) radius in MIRI/F1500W and in HST/F160W \citep{vanderwel14}. $^{\dagger}$ASPECS-15 is an X-ray AGN.
\end{minipage}
\end{table*}

\subsection{Sample}
\label{sec:sample}

PAHSPECS targets all (5) galaxies, without any prior selection, detected at $z\sim1.1$ in the 1 and 3\,mm maps of the deep, blind ASPECS survey \citep{Walter2016,Decarli2019b,Aravena2020}, tracing CO(2-1) and dust emission, respectively. Even if apparently small, this is an unbiased and complete ALMA-selected sample at $z\sim1.1$ -- a redshift that allows us to target prominent PAH features within the MIRI MRS spectral coverage. In other words, the targets represent the full sample of dust- and/or CO-detected galaxies in the HUDF at $z\sim1.1$, complete down to a molecular gas and dust mass of $M_{\mathrm{mol}} > 7 \times 10^9\,M_{\odot}$ and $M_{\mathrm{dust}} > 6\times 10^7M_{\odot}$ (at 5$\sigma$; \citealt{Boogaard2019}), respectively. The selection of these targets demonstrates the diversity in the population of gas-rich, star-forming galaxies at $z\sim 1$ (Table~\ref{tab:pahspecs_sample_props}). The galaxies are diverse in (cold) dust-to-gas ratio (as mapped by the dust to CO luminosity) with molecular gas depletion timescales of the order of 1\,Gyr and a molecular gas content comparable to the stellar mass \citep[for additional molecular gas and dust related properties, see][]{Boogaard2020,Aravena2020,GL2020}. 

\begin{figure}[ht]
        \centering
        \includegraphics[width=.9\columnwidth]{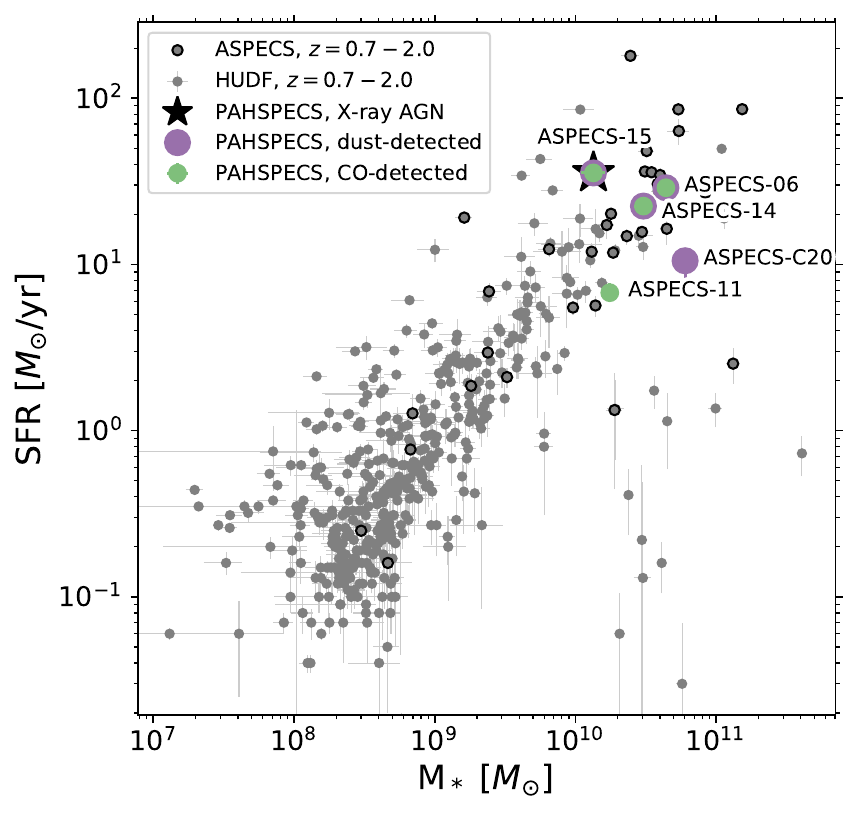} 
    \caption{SFR versus stellar mass of PAHSPECS galaxies in comparison to the HUDF sample \citep{Boogaard2020}. The HUDF sample contains every galaxy within the redshift range of $0.7-2.0$ with F444W flux density $>$\,100\,nJy, fit in the same way as the PAHSPECS galaxies. PAHSPECS galaxies, while sampling the more massive/star-forming end, lie on the general main-sequence trend of galaxies in the HUDF. Among the PAHSPECS sample, there is an X-ray AGN (black star). Four targets are detected in CO(3-2) (green circles), and four in 1\,mm dust continuum (purple circles). The CO/dust continuum detected sources of the ASPECS survey within the redshift range of $0.7-2.0$ are shown with black edges.
    }
    \label{fig:MS}
\end{figure}

Furthermore, the sample is diverse in nuclear activity.  Notably, ASPECS-15 has been detected in X-rays and identified as showing signs of AGN activity \citep{Luo2017}. Their optical F160W effective radii vary from $\sim1.8$\,kpc to $>5$\,kpc while their mid-IR F1500W effective radii are more coherent (Table~\ref{tab:pahspecs_sample_props}). They have moderate to high stellar masses ($M_{\mathrm{\star}} \gtrsim 10^{10}\,M_{\odot}$; Table~\ref{tab:pahspecs_sample_props}) and while the position of the galaxies with respect to the main-sequence varies (Figure~\ref{fig:MS}), all sources lie within $\sim \pm 3$ times the scatter. Based on the [O{\sc ii}]$\lambda3727$/[Ne{\sc iii}]$\lambda3869$ ratios of the sample measured from MUSE spectroscopy, their gas-phase metallicity is around solar \citep{Boogaard2019}, in agreement with their stellar mass. By probing the PAH emission, PAHSPECS explores how variations in the ISM properties can arise in a purely CO and cold dust-selected sample of galaxies at cosmic noon (see Section~\ref{sec:Results}).

\subsection{Survey strategy and data reduction}
\label{sec:survey}

\subsubsection{MRS observations}
\label{sec:MRS-strategy}

Observations of the five sample galaxies were carried out on November 12-14, 2024 with the integral field unit (IFU) of the mid-infrared instrument \citep[MIRI;][]{Wright2023,Argyriou2023} onboard JWST, as part of the Cycle 3 program 5279 entitled ``PAHSPECS: A Comprehensive Study of PAHs at cosmic noon'' (PI: Shivaei, co-PIs: D{\'i}az-Santos, Boogaard).

The goal of the PAHSPECS program is to detect the main PAH features at rest-frame 3.3, 6.2, 7.7, and 11.3\,$\mu$m. At $z\sim 1.1$, these features are at observed $\sim 6-27\,\mu$m, therefore all four channels of MRS are used to capture the full set of emission features, underlying continuum, and absorption features.

The program total allocated time was 52.6 hours. Exposure times were estimated based on the best-fit SEDs of each individual galaxy to the NIRCam+MIRI photometry from 0.9 to 25\,$\mu$m as input spectrum in the ETC. To increase efficiency, only gratings that include the PAH features at 3.3 and 11.3\,$\mu$m (gratings C and B, respectively) required long exposure times. Therefore, grating A, which contains the 6.2\,$\mu$m feature only, has 1/4th of the exposure time of gratings B and C. All sources have the same exposure times as they are similar in surface brightness, except for source 3mm.11, which is the faintest one and has 1.4 times longer exposures. 

To reach to the desired SNR, we chose 15-20 groups per integration and different integrations for each of the 3 grating settings. We adopted a 4-point extended source dither pattern as the best way to ensure that observations achieve the optimal spatial sampling. We adopted SLOWR1 readout mode as the commonly-used FASTR1 readout mode would have resulted in exceeding the allowed data rate limit.   

Dedicated background observations are required for relatively faint sources, such as in this program, to ensure best performance. As the targeted galaxies have a variety of morphologies, finding a portion of the field of view where a local background can be measured, especially in the most narrow Channels 1 and 2, was considered not practical. Therefore, we took dedicated background observations, which we designed to be as efficient as possible. We took a dedicated background with the same detector readout parameters as our observations (i.e., the same number of groups and integrations) but only 2-point dither pattern (instead of 4). The dither provides robustness against artifacts and improves on the PSF sampling compared to no dithering. This method was adopted to facilitate the "master background" subtraction in the JWST pipeline, which in theory increases the SNR of the background by averaging it over the entire field. However, while we tried this method in data reduction, we decided to use a pixel-by-pixel background subtraction at the end (see Section~\ref{sec:MRS-datareduc}).

Our five targets were closely-spaced within 1-2 arcmin of each other, so we grouped two of them together, saving telescope time by avoiding target acquisition twice. We also requested only one dedicated background for one of the targets (ASPECS-6) as part of an uninterruptible sequence and set a special requirement to observe the other targets within 1 day of the observations. We used a single background observation for all science targets to balance between the most efficient and optimal quality data reduction.

\subsubsection{MRS data reduction}
\label{sec:MRS-datareduc}

The JWST/MIRI-MRS data were reduced using the official JWST Science Calibration Pipeline (v1.17.0; \citealt{Bushouse2025}) with the Calibration Reference Data System (CRDS) context jwst\_1321.pmap, following the standard three-stage processing scheme. In Stage 1, detector-level calibrations were applied to the uncalibrated exposures. We enabled cosmic ray detection, while all other steps followed the default pipeline implementation. In Stage 2, the spectroscopic calibration was applied, including WCS assignment, fringing correction, flat-fielding, and photometric calibration. We also applied an additional targeted cosmic-ray mitigation step to reduce residual artifacts from cosmic ray showers\footnote{This step was later added to the official pipeline: https://jwst-pipeline.readthedocs.io/en/latest/jwst/straylight/main.html\#algorithm-residual-shower-correction}, and enabled residual fringing correction. At this stage, background subtraction was carried out using our dedicated off-source observation. In particular, the background field associated with ASPECS6 was used for all sources (see Section~\ref{sec:MRS-strategy}), and the background was subtracted on a pixel-by-pixel basis, as discussed above. Stage 3 combined the calibrated exposures into flux-calibrated 3D IFU data cubes, with outlier rejection and bad pixel correction enabled.

The final products consist of 12 data cubes corresponding to the four MIRI/MRS channels and their SHORT, MEDIUM, and LONG sub-bands, providing continuous wavelength coverage from 4.9 to 27.9\,$\mu$m. For the final data products, IFU-aligned cubes were adopted, preserving the detector orientation and facilitating the correction of detector-aligned artifacts. A destriping procedure was applied to mitigate residual row-correlated features, resulting in a systematic reduction of the noise level. This approach, also used by the TEMPLATES team for their data-processing \citep{Rigby2025}, was independently validated through synthetic photometry, which showed improved agreement between the spectroscopic and imaging fluxes. For further details on the data reduction procedure, we refer the reader to \cite{Lofaro2026}. 

\begin{figure*}[ht]
        \centering
        \includegraphics[width=.8\textwidth]{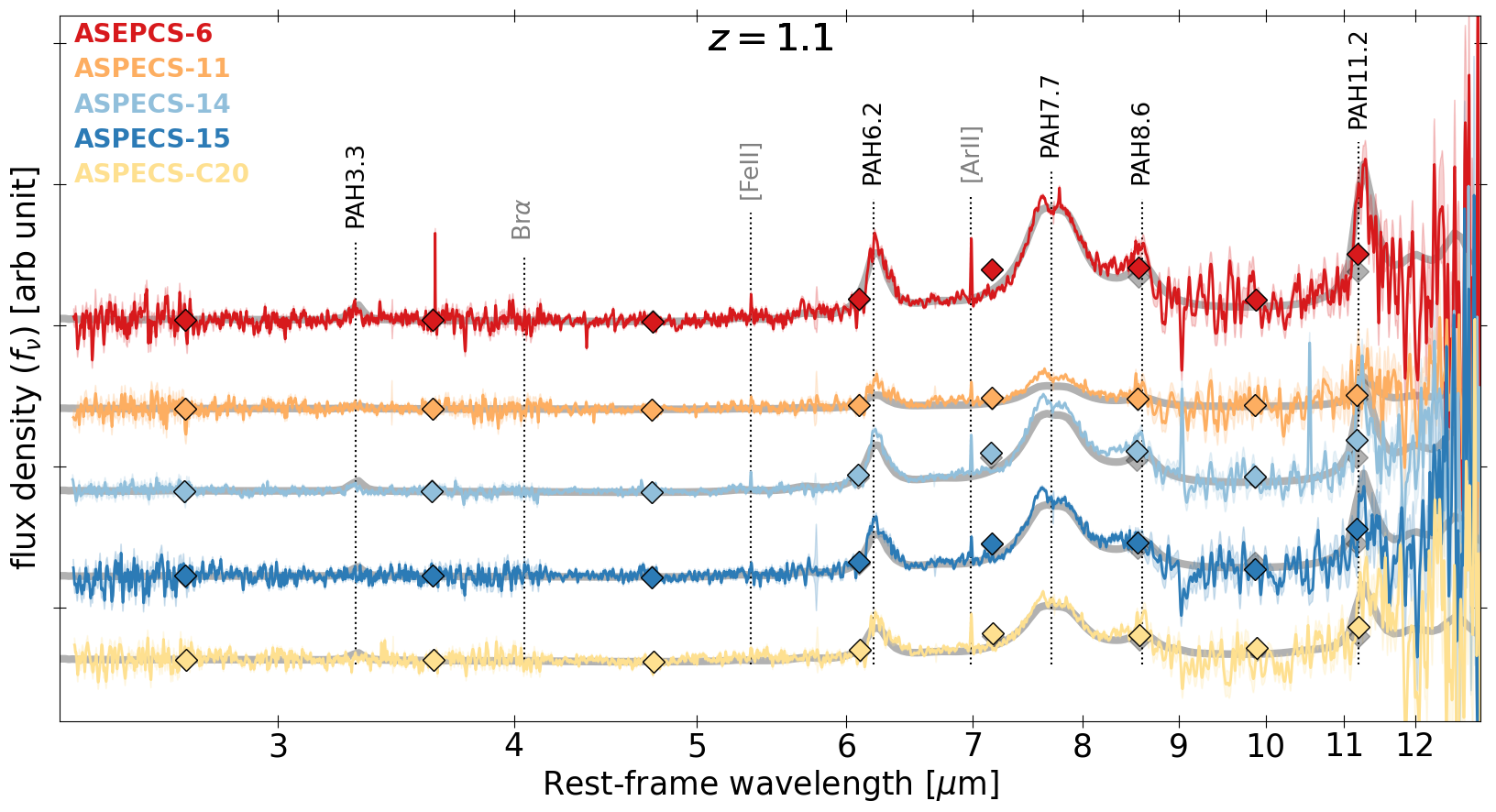} \\
        \includegraphics[width=.18\textwidth]{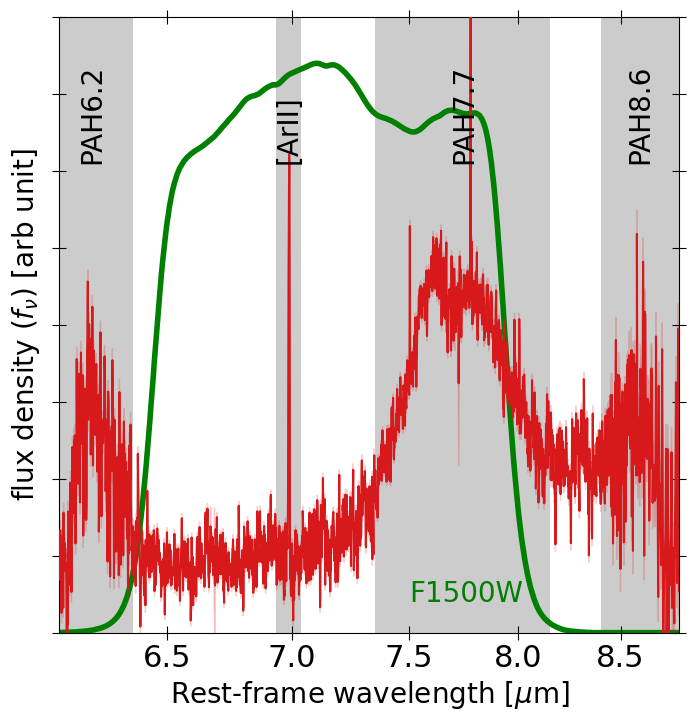} \quad
        \includegraphics[width=.18\textwidth]{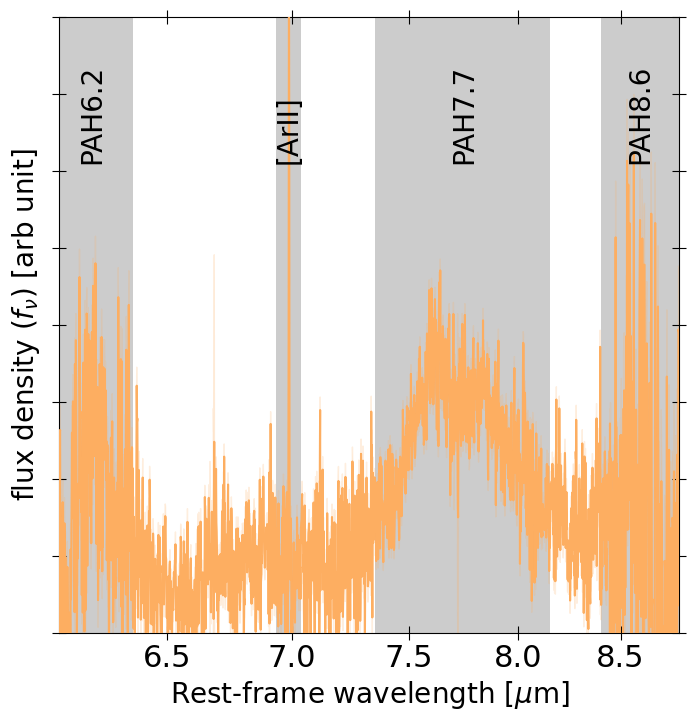} \quad
        \includegraphics[width=.18\textwidth]{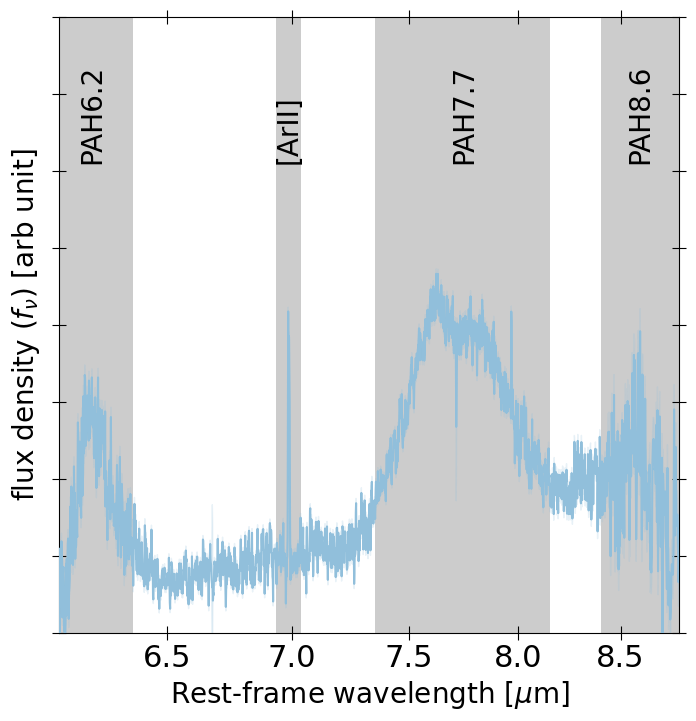} \quad
        \includegraphics[width=.18\textwidth]{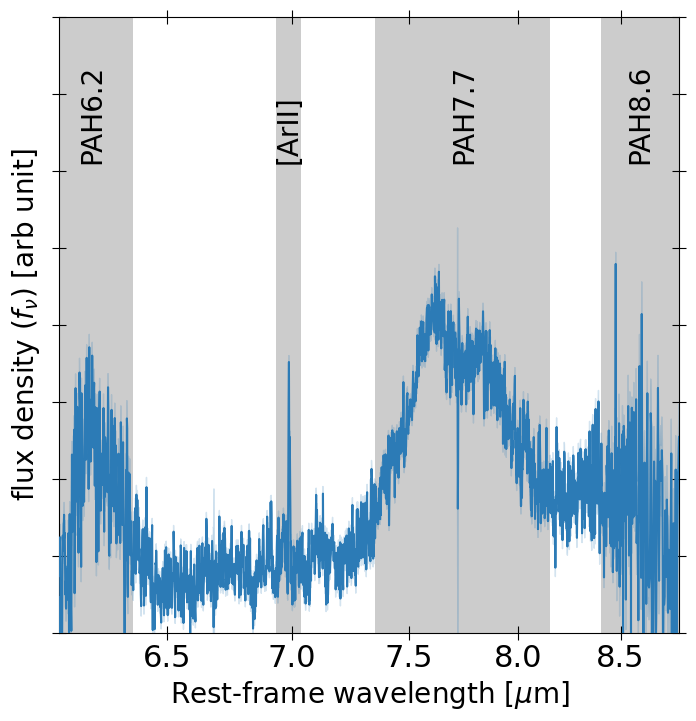} \quad
        \includegraphics[width=.18\textwidth]{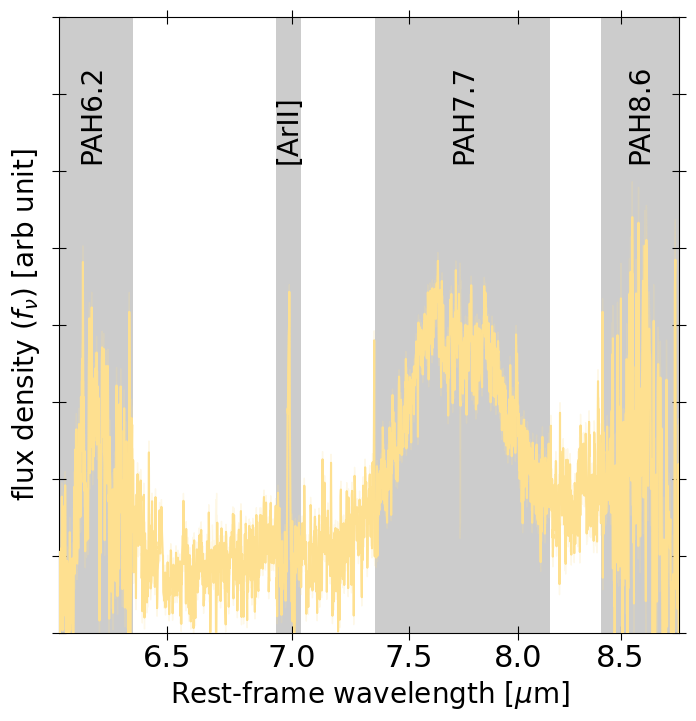}
        
    \caption{Top: MRS spectra of the five PAHSPECS galaxies, with MIRI photometry (color diamonds) and best-fit SEDs (thick grey line) overplotted. Each galaxy is shown with a different color and the spectra are shifted arbitrarily in vertical direction and smoothened by a Gaussian kernel with $\sigma=4$ for better visualization. The MIRI photometry from 5.6 to 25$\,\mu$m is shown with the same colors. The best-fit SED to the photometry only is shown with thick grey line for each galaxy.
    Bottom: The native resolution MRS spectra is shown in the region of highest SNR at $\lambda_{\mathrm{rest}}\sim$\,6--9\,$\mu$m. [Ar{\sc ii}] at 6.98\,$\mu$m is detected in all galaxies. For the 7.7\,$\mu$m feature, the 7.42, 7.60, and 7.85\,$\mu$m components are shown. The MIRI F1500W filter transmission curve is shown for reference in the panel of ASPECS-6 with a green line, as it is used in Section~\ref{sec:photvsspec} to perform synthetic photometry on the MRS spectra}.
    
    \label{fig:spectra}
\end{figure*}

As detailed in \cite{Lofaro2026}, we extracted 1D spectra from the MRS cubes using wavelength-dependent circular apertures centered on each galaxy. Because the sources are faint in the MRS data, the aperture sizes were empirically determined from the higher-S/N multi-band MIRI imaging \citep[from SMILES;][]{Alberts2024, Rieke2024}: at each wavelength, the radius that maximized the S/N based on the source curve of growth, typically enclosing 60\% of the total flux, was selected, and then fitted a linear relation between aperture radius and wavelength. An aperture correction was subsequently applied to recover the total flux. The spectra of PAHSPECS galaxies are shown in Figure~\ref{fig:spectra}.

\subsubsection{Simultaneous F1500W Imaging} \label{sec:miri-imaging}

PAHSPECS also has simultaneous MIRI imaging to improve the astrometric solution of individual MRS exposures as our extended sources did not require prior target acquisition observations. The simultaneous imaging was done in F1500W. Given the very long exposure times for each simultaneous imaging, specific PA range was requested to overlap the imaging on deep MIDIS and MIDI-RED 5.6, 7.7 and 10\,$\mu$m footprints for maximum science return (JWST Prog. IDs 1283 and 6511, PI: G. \"{O}stlin; \citealt{ostlin25}; see \citealt{Boogaard2024} for an analysis of the ASPECS and PAHSPECS sources from this data). Additionally, they were positioned to also cover the MRS sources (cutouts shown in Figure~\ref{fig:seds} and \ref{fig:seds2}). With the six pointings combined (five science and one background), the exposure time of the F1500W images are between 15 to 25 hours, making these the deepest existing mid-IR images of the sky beyond 10\,$\mu$m that opens opportunities to exciting discoveries and robust constraints on the SEDs of high redshift galaxies \citep[e.g.,][]{rinaldi25, perez-gonzalez26, FA2026}.

\subsubsection{F1500W imaging data reduction}
\label{sec:miri-dr}

The MIRI imaging data was reduced with the Rainbow JWST pipeline, which is based on the official {\sc jwst} package with the addition of some bespoke offline procedures to deal with the inhomogeneous MIRI imager thermal background \citep[see][for discussion on the imager background]{dicken24}. The main steps are described in \citet{perez-gonzalez24} and \citet{perez-gonzalez26} (see also \citealt{Alberts2024, ostlin25}). Briefly, after running stage 1 and 2 of the official {\sc jwst} pipeline, version 1.16.0 and pmap\_1202, the background of each integration is homogenized by subtracting a super-background frame. This super-background frame comes from a median stack of all the rest of frames in the dataset, for which the sources have been masked. The source masking needs a detection execution, so the method is iterative: we construct a mosaic with all the data, detect sources and build a mask, build the super-background taking into account those masks, apply it to the stage 2 images, and re-mosaic. Two iterations are enough to reach optimal results. This method typically helps to improve the depth of the final mosaics by 0.3-0.6~mag (see references above). The 5$\sigma$ depth, measured from aperture photometry using 0.5 arcsec radius apertures, varies between 26.62 and 25.61 mag (AB), with a mean of 26.05 mag (AB) across the field \citep{FA2026}.

\section{SED fitting} \label{sec:sed}\
\label{sec:SEDfitting}

Stellar population and dust inferences are performed using the \texttt{Prospector} Bayesian SED fitting code \citep{Leja2017, Leja2019, Johnson2021} closely following the choices outlined in \cite{Shivaei2024a} and \cite{Shivaei2024b}. The SED fitting is done to the rest-frame UV to IR photometric data alone and redshifts are fixed to the spectroscopic redshifts. We use the HST and JWST/NIRCam photometry from JADES DR5 \citep{Johnson2026,Robertson2026} including the JWST/MIRI photometry from SMILES \citep{Alberts2024,Rieke2024}. We use the Herschel photometry from \citet{Elbaz2011} and ALMA 1.2 and 3\,mm photometry from ASPECS \citep{Boogaard2020}. Notably, we include upper limits on the Herschel and ALMA bands for all sources based on the respective depths of the catalogs. The best-fit SEDs of the PAHSPECS sample are shown in Figure~\ref{fig:seds}. Throughout this work, we report the surviving stellar mass, which accounts for mass loss during stellar evolution (e.g., AGB winds and supernovae), rather than the total formed stellar mass. We assume a \cite{Chabrier2003} IMF. 

\texttt{Prospector} uses energy balance to connect the UV-optical light to the dust emission in IR. Within \texttt{Prospector}, we adopted the Flexible Stellar Population Synthesis \citep[FSPS;][]{Conroy2013} code to model the main physical properties of galaxies.
For building the stellar and nebular models, we adopt the MESA Isochrones and Stellar Tracks \citep[MIST;][]{Choi2016}, a set of stellar evolution models that provide isochrones and tracks used in population synthesis, and the Medium-resolution Isaac Newton Telescope Library of Empirical Spectra \citep[MILES;][]{FB2011} spectral library, an empirical stellar spectral library that supplies observed spectra of stars across a range of parameters. We assume a delayed-$\tau$ star formation history, with a uniform prior on the stellar age ranging from 1\,Myr to the age of the Universe at the redshift of each galaxy.  

Nebular emission is modeled with the CLOUDY \citep{Ferland2017} grid presented in \cite{Byler2017}, adopting logarithmic priors for the gas-phase metallicity (from $-2.0$ to 0.5 relative to solar) and the ionization parameter (from $-4$ to $-1$). The stellar metallicity is treated as a free parameter, with a flat prior over $\log(Z/Z_{\odot}) = -2.5$ to 0.19. Absorption by the intergalactic medium is modeled following \cite{Madau1995}, with the optical depth normalization allowed to vary, assuming a Gaussian prior centered at 1.0 with $\sigma = 0.3$. 

We adopt a modified dust attenuation curve following \cite{Shivaei2025}, in which the UV--optical slope and the UV bump amplitude are treated as independent parameters, as in \cite{Noll2009, Salim2018}. The slope is parameterized as a multiplicative modification of the \cite{Calzetti2000} curve, with a flat prior in the range $-0.6$ to 0.3.

For the dust emission, we adopt the dust model of \cite{Draine2007}, which assumes a mixture of amorphous silicate and carbonaceous grains. In these models, dust is heated by a radiation field with the spectral shape of the local interstellar radiation field \citep{Mathis1983}, scaled by a dimensionless factor $U$. The model has three free parameters: ${\rm q_{PAH}}$, the mass fraction of dust in PAH grains containing fewer than $10^3$ carbon atoms; $U_{\rm min}$, the minimum radiation field intensity heating the diffuse ISM; and $\gamma$, the fraction of dust mass exposed to a power-law distribution of starlight intensities between $U_{\rm min}$ and $U_{\rm max}$ (the remaining fraction $1-\gamma$ being heated at $U_{\rm min}$). We adopt uniform priors in the ranges $U_{\rm min}=0.1$–15, $\gamma=0.001$–0.15, and ${\rm q_{PAH}}=0.4$–4.6, corresponding to the parameter space over which the \cite{Draine2007} models are defined.
The PAH luminosities estimated from the best-fit SED models are calculated by integrating the features within specific wavelength ranges around each PAH \citep{Draine2021} after performing a local continuum subtraction using a polynomial of order one (see Figure~\ref{fig:totclip}).

\section{Results}
\label{sec:Results}

In Figure~\ref{fig:spectra}, we show the final spectra of five PAHSPECS galaxies (see Section~\ref{sec:MRS-datareduc} for spectral extraction details), arbitrarily shifted in vertical direction for visual purposes. 
The spectra in the top panel are smoothed with a Gaussian kernel with a width of 4$\sigma$ to better highlight the broad PAH features. We detect the PAH6.2, 7.7 and 11.3 in all sources, while the PAH3.3 is detected in two galaxies, ASPECS-6 and ASPECS-14. We note that \cite{Donnan2026b} also recovers the feature in ASPECS-15, using a spatially resolved, forward-modeling approach. The bottom panels in Figure~\ref{fig:spectra} show the full/native resolution spectra in the $\sim$\,6--9\,$\mu$m wavelength range. The [Ar$\rm\,II$] line at 6.98\,$\mu$m is detected in all sources, and the SNR is sufficient to identify all the sub-components of the PAH7.7 complex at 7.42, 7.60, and 7.85$\,\mu$m.

\subsection{Recovering PAHs in spectroscopy versus photometry}
\label{sec:photvsspec}

In addition to the MRS spectroscopy, Figure~\ref{fig:spectra} also shows the MIRI multi-band photometric data from the SMILES survey \citep{Rieke2024, Alberts2024} and the best-fit SED to this photometry (Section~\ref{sec:sed}). There is a slight offset between the broad-band photometry and the MRS spectra, which likely stems from uncertainties associated to aperture definition and corrections in photometry. Particularly, in the case of ASPECS-15 the field is very crowded with multiple galaxy projections in the line of sight, which makes the aperture definition and correction challenging. To account for these offsets, we calculated an average scaling factor based on synthetic photometry performed to the MRS spectra and the SMILES imaging photometry of each source, from the F560W to the F1500W filters. We refer the reader to \cite{Boogaard2026} for further details. The largest correction factors are those of ASPECS-15 (photometry/MRS $\simeq$\,1.87) and ASPECS-11 ($\simeq$\,1.68). ASPECS-6, 14 and C-20 have correction factors of $\simeq$\,1.08, 1.38 and 1.38, respectively. The correction factors are applied to the MRS spectra shown in Figure~\ref{fig:spectra} as well as to all the MRS-derived PAH fluxes and luminosities used in the rest of the paper.

In this section, we compare the PAH fluxes inferred from the SED fitting to the photometric data with those obtained from spectral decomposition fitting to the MIRI IFU data. We first describe the results inferred from the integrated PAH fluxes (Section~\ref{sec:intfluxes}) and then present the spatially resolved spectral PAH7.7 maps, which we compare to our MIRI F1500W images (Section~\ref{sec:resolved-maps}).

\subsubsection{Integrated PAH fluxes}
\label{sec:intfluxes}

\begin{figure}[ht!]
        \centering
        \includegraphics[width=\columnwidth]{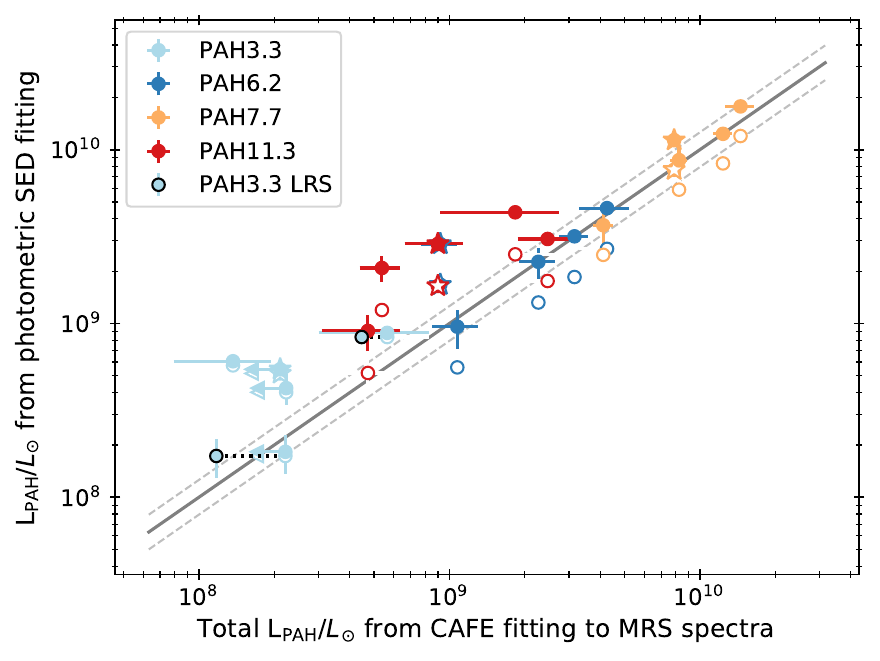} 
    \caption{Comparison of PAH luminosities obtained from the MIRI/MRS spectra (Figure~\ref{fig:spectra}) and those obtained from the best-fit SEDs to the MIRI multi-band photometry for the 5 PAHSPECS galaxies (Figures~\ref{fig:seds}, \ref{fig:seds2}). The data for each PAH feature are shown as open circles, with colors representing different PAH features. All MRS luminosities are scaled up by the photometry/MRS factor described in Section~\ref{sec:photvsspec} \citep{Boogaard2026}. However, these data-points do not include the corrections to the PAH luminosities obtained from the photometric SED fits, which only account for \textit{clipped} luminosities (see Section~\ref{sec:intfluxes}). When the clipped-to-total factors are applied, we obtain the solid circles. The only X-ray AGN in the sample, ASPECS-15, is shown with a star symbol. The gray solid line shows the 1--1 ratio and the dotted lines show $\pm$\,0.1\,dex ($\sim$\,25\%) intervals around the line. The three undetected PAH3.3\,$\mu$m, 1$\sigma$ upper limits are shown with arrows. Symbols with black edges denote PAH3.3 measurements obtained from MIRI/LRS spectroscopy from the GTO program 1533 (PI: G.~{\"O}stlin, \citealt{Kendrew2026}) for the same galaxies. The MRS and LRS datapoints of the same galaxy are connected with a dotted line. We can see that the 6.2 and 7.7\,$\mu$m PAHs are recovered very well by the photometric SED fitting methodology once all correction factors are applied, while the 11.3 and 3.3\,$\mu$m PAHs are on average overestimated.}
    \label{fig:PAHflux}
\end{figure}

The open symbols in Figure~\ref{fig:PAHflux} show the comparison between the PAH luminosities derived from the MRS spectra using the Continuum And Feature Extraction code \texttt{CAFE}\footnote{\href{https://github.com/GOALS-survey/CAFE}{https://github.com/GOALS-survey/CAFE}} \citep{DS2025} and those obtained from the best-fit SED to the photometric data.

\texttt{CAFE} decomposes JWST IFU (MIRI-MRS and NIRSpec-IFU) spectra into multiple components, including stellar emission, dust continuum, PAH features, and emission lines---with PAH bands modeled using Drude profiles--, all subject to dust attenuation \citep{Marshall2007}. As a consequence, PAH luminosities obtained from \texttt{CAFE} fitting refer to \textit{total} luminosities, and include all the power associated to the PAH wings. Throughout this work all the \texttt{CAFE}-derived luminosities are not corrected for extinction, as mid-infrared dust attenuation is expected to be small in these main-sequence galaxies.\footnote{We note that the foreground extinction in the HUDF is negligible.} We refer to \cite{Lofaro2026} for the details on the \texttt{CAFE} fitting procedure to the MRS spectra and the discussion on the integrated PAH properties of the sample. We only note here that the integrated MRS spectra are fitted together with MIRI photometry in the $\lambda_{\rm obs}$\,$\simeq$\,2--3 and 15--25\,$\mu$m wavelength ranges, to aid the fitting and ensure an accurate calculation of the underlying continuum under the PAH features.

\begin{figure}[t]
        \centering
        \includegraphics[width=.9\columnwidth]{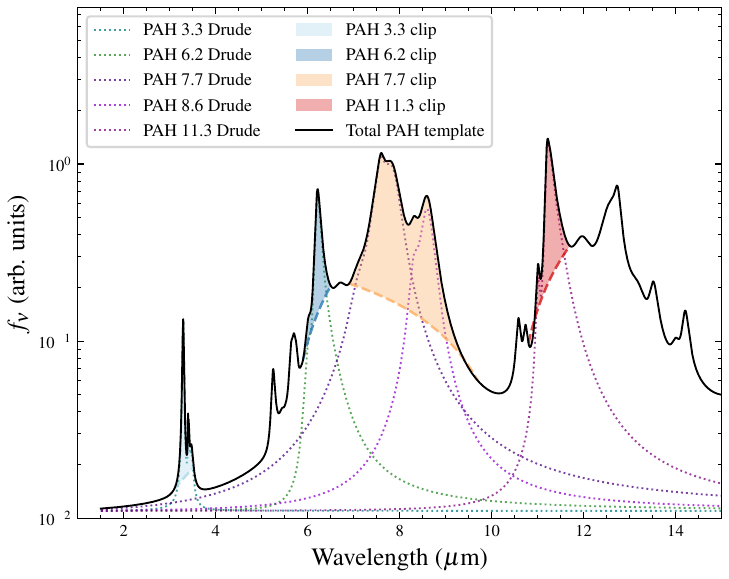} 
    \caption{A pure PAH template spectrum showing main PAH features from rest-frame $\sim$\,2 to $\sim$\,15$\,\mu$m. The template is the average of 12 star-forming regions in local LIRGs observed with JWST and continuum/line subtracted \citep{Rigopoulou2024}. The colored regions show the ``clipped'' PAH integrated areas after accounting for the local continua (dashed lines) as defined by \cite{Draine2021}. On the other hand, spectral decomposition tools such as \texttt{CAFE} \citep{DS2025}, \texttt{PAHFIT} \citep{Smith2007}, or \texttt{SPIRIT} \cite{Donnan2024}, use Drude profiles (dotted lines) to fit the PAH features to account for their total emission. 
    The wings of the PAH features extend beyond the clipped ranges and contribute non-negligibly to the total (true) fluxes of the features. In Table~\ref{tab:totalclip}, we report the total-to-clipped flux ratio for each PAH (or PAH combination), defined as the ratio between the luminosity integrated over the full Drude profile and that integrated over the corresponding colored (clipped) region.
    }
    \label{fig:totclip}
\end{figure}

\begin{table}[]
\centering
\footnotesize
\caption{Total (Drude) to Clipped PAH flux ratios}
\label{tab:totalclip}
\begin{tabular}{lcc}
\hline\hline
PAH band  &  $\lambda$ range  &  Total/clipped ratio \\
 ($\mu$m)      &   ($\mu$m)   &   \\
\hline
3.3  &     3.09--3.52  & 1.058  \\
3.3+3.4 &  3.09--3.52  & 1.304  \\
6.2 &      5.90--6.50    & 1.715  \\
7.7 &      6.90--9.70    & 1.186  \\
7.7+8.6 &  6.90--9.70    & 1.480  \\
8.6 &    6.90--9.70    & 0.293  \\
11.3  &    10.80--11.70  & 1.745  \\
17  &    15.50--18.50  & 1.464  \\
\hline
\end{tabular}
\begin{minipage}{0.45\textwidth}
\vspace{0.25cm}
\footnotesize
\textbf{Notes.} Columns: (1) PAH band(s) included in the ratio; (2) Wavelength ranges used to integrate PAH bands directly from spectra (after continuum subtraction), as defined in the clipping methodology from \cite{Draine2021}; (3) Flux ratios between the full Drude profile used to model the \textit{total} PAH emission by spectral decomposition tools and the \textit{clip} methodology (see Figure~\ref{fig:totclip}).
\end{minipage}
\end{table}

The PAH luminosities estimated from the best-fit SED models to the broad-band photometry are calculated following the \textit{clipping} methodology of \citep{Draine2021}. In brief, wavelength ranges are defined around the PAHs of interest (see Table~\ref{tab:totalclip}) and used to integrate the emission of the features directly on the spectra, after a linear continuum subtraction is performed in $\nu f_\nu$--log($\lambda$) space using as anchors the boundaries of the clipping ranges. These clipped regions are shown as shaded colored areas in Figure~\ref{fig:totclip}, and for comparison, modeled PAH emissions using a Drude profile are overplotted with dotted lines. As shown, the extracted clipped luminosities from the \citep{Draine2021} regions do not account for the most extended wings of the PAHs, contrary to the \texttt{CAFE} decomposition that uses Drude profiles. However, it is possible to calculate a correction factor to go from the \textit{clipped} luminosity to \textit{total} luminosity by using a scale-free PAH template and the clipping ranges defined by \cite{Draine2021} as shown in Figure~\ref{fig:totclip}. Table~\ref{tab:totalclip} presents a summary of the total-to-clipped flux ratio for every PAH (or combination of PAHs in a given region). We use these ratios to correct the luminosities of the PAHs derived from the best-fit photometric SEDs in Figure~\ref{fig:PAHflux} (solid circles), so they can be fairly compared to those obtained from the \texttt{CAFE} decomposition of the MRS spectra.

We also note that the wavelength range used to integrate the PAH7.7 complex on the best-fit SEDs encompasses the PAH8.6 feature. For this reason, in Figure~\ref{fig:PAHflux} we also plot the sum of both features when showing the luminosities derived from the \texttt{CAFE} fit. Finally, the wavelength range used to integrate PAH3.3 also includes both the PAH3.3 and PAH3.4 features in the \textit{clipped} method. In this case, we adopted a 10\% contribution factor of the PAH3.4 based on \cite{McKinney2026} and subtract it from the \textit{clipped} PAH3.3 luminosity obtained from the best-fit SED. Triangles indicate 1$\sigma$ upper limits on the PAH3.3 fluxes for non-detections.

After applying the \textit{clipped}-to-\textit{total} luminosity factor (Table~\ref{tab:totalclip}) to the best-fit SED-derived luminosities, Figure~\ref{fig:PAHflux} shows (solid symbols) that the PAH7.7+8.6 luminosity is recovered remarkably well from the photometric SED fit for the PAHSPECS star-forming galaxies. As described above, this result is encouraging and suggests that it is possible to use MIRI multi-band photometry alone to derive accurate PAH7.7 complex luminosities ($<$\,20\% difference). Similarly, the PAH6.2 feature is also well recovered, with values consistent with those derived from the MRS spectra to within $\sim$\,25\%; with the exception of ASPECS-15, the X-ray AGN in our sample, which shows an excess of a factor of $\sim$\,3 above the 1--1 relation. 

The recovery of the PAH11.3 and PAH3.3 features from photometry and SED models is less reliable, with the models often overestimating their luminosities by more than a factor of $\sim$\,2. In particular, the \cite{Draine2007} SED models used by \texttt{Prospector} significantly over-predicts the PAH3.3 luminosity for the majority of the sample, especially for faint sources, with differences that can be as large as a factor of $\sim$\,5, highlighting that this feature is substantially fainter than the model predicted value in galaxies at cosmic noon.

\subsubsection{Resolved PAH maps}
\label{sec:resolved-maps}

\begin{figure*}[ht!]
        \centering
        \includegraphics[width=\textwidth]{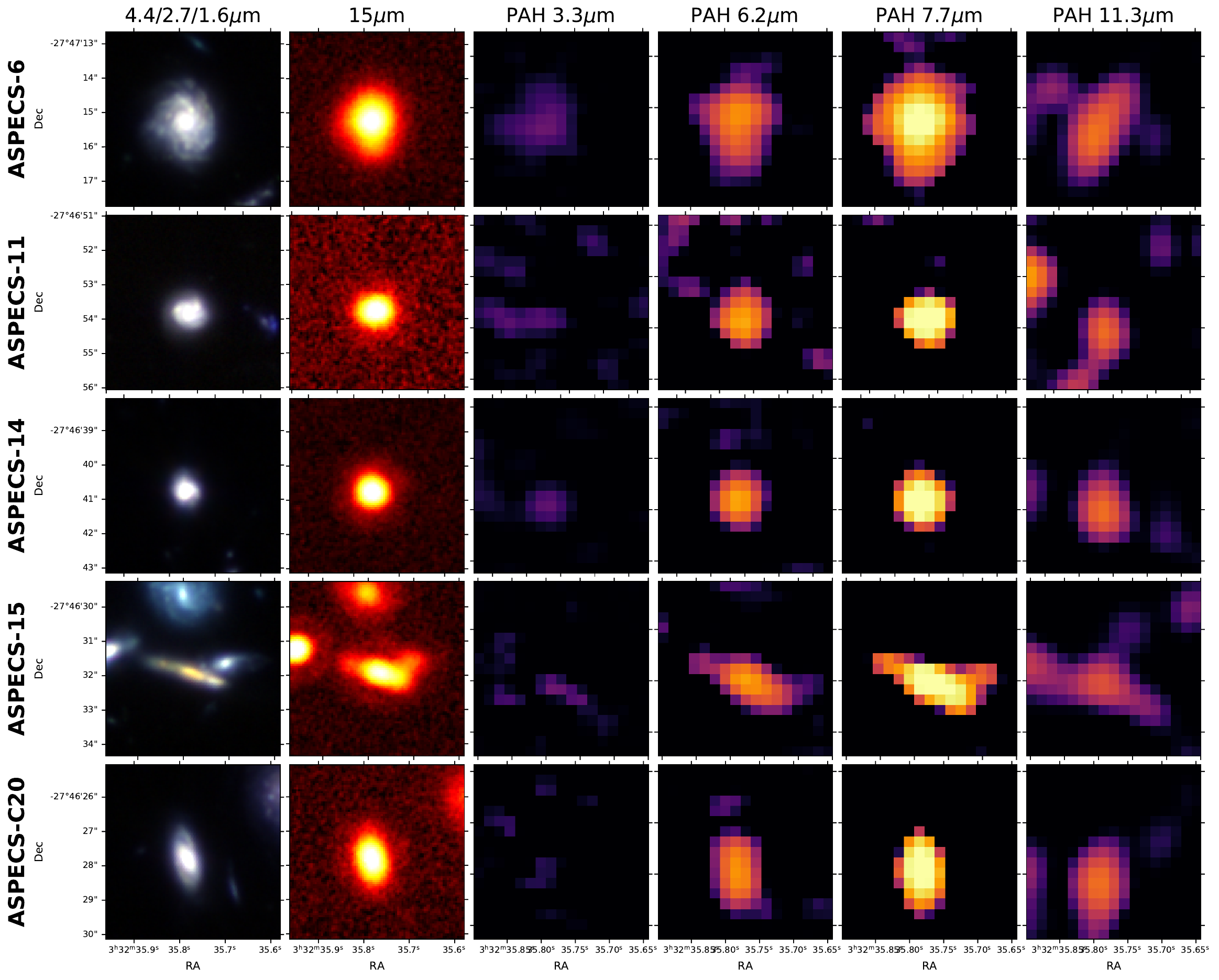}
    \caption{ 
    Multi-wavelength cutouts of PAHSPECS galaxies. , from left to right, RGB from JWST/NIRCam F444W, JWST/NIRCam F277W, and HST/WFC3 F160W filters (JADES), JWST/MIRI F1500W (Section~\ref{sec:miri-imaging}), and JWST/MRS continuum-subtracted emission line maps of PAH 3.3, 6.2, 7.7, and 11.3\,$\mu$m features \citep{Donnan2026b}. The four PAH maps in each row share the same color scale so that the relative strengths of the different PAH features can be compared within each source (the color scale is not shared between rows). 
    }
    \label{fig:PAH-mosaic}
\end{figure*}

\begin{figure*}[ht!]
        \centering
        \includegraphics[width=\textwidth]{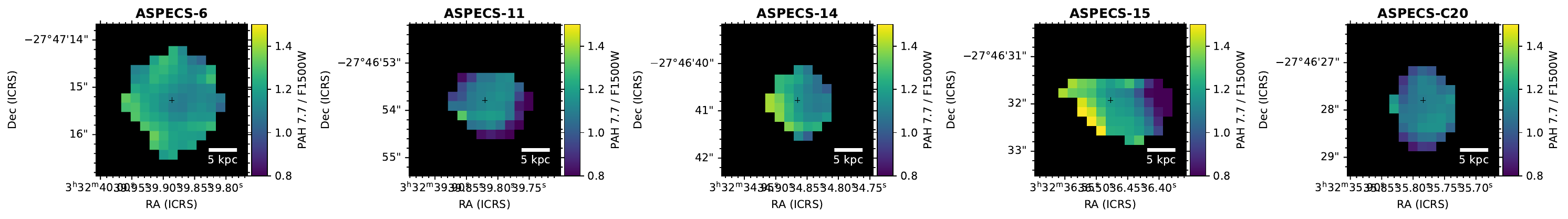} \\
        \includegraphics[width=\textwidth]{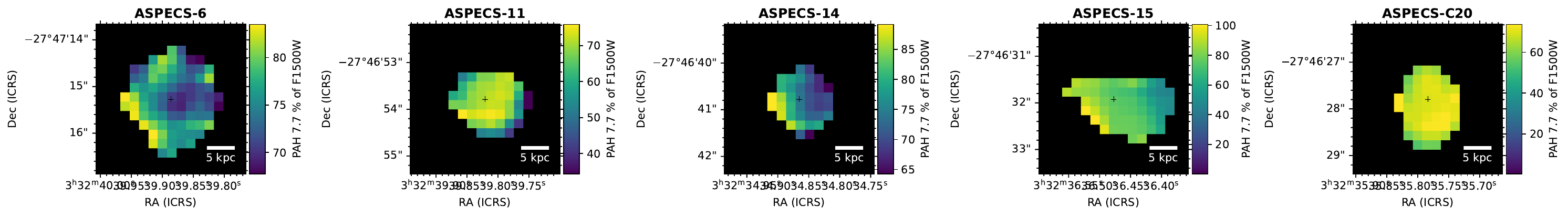} \\
        \includegraphics[width=\textwidth]{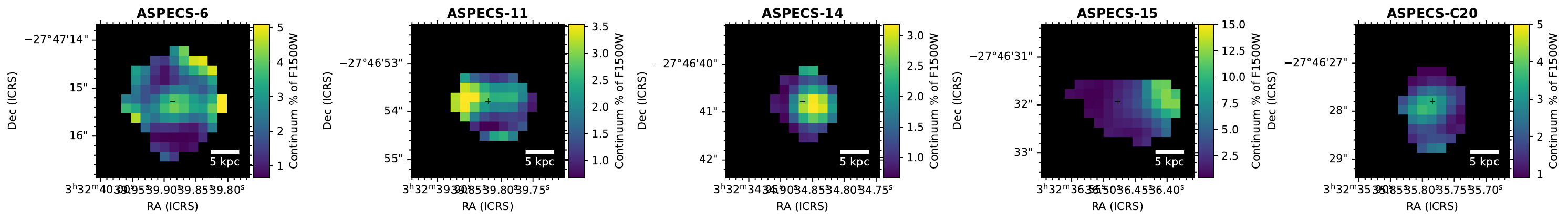} 
    \caption{Ratio maps involving the PAH7.7 feature and the F1500W synthetic photometry applied to the MRS cubes for ASPECS galaxies. The first row of panels shows the maps of the total PAH7.7 feature to the synthetic F1500W photometry. Ratios above 1 are possible because the wings of the PAH feature extend well beyond the filter (see text for details and \citealp{Donnan2026b}). The second row of panels is the fraction of PAH7.7 feature within the F1500W filter to the synthetic F1500W photometry. The third row of panels is the same as the second but for the fraction of continuum emission within the F1500W to the synthetic F1500W photometry. 
    }
    \label{fig:PAH-maps}
\end{figure*}

Figure~\ref{fig:PAH-mosaic} presents the maps of the PAHSPECS galaxies in four PAH bands. 
The PAH maps are constructed by forward modeling of the MRS data cubes as described in \cite{Donnan2026b}. Briefly, each wavelength range (or filter) containing a PAH feature is modeled independently as a Drude profile plus an underlying slope that accounts for the continuum emission, producing a model cube that is fitted to the data spaxel by spaxel until the optimization converges after enforcing a degree of regularization among contiguous spaxels (see more details in \citealt{Donnan2026b}). The PAH features at 6.2, 7.7, and 11.3\,$\mu$m show clear detections in their spatially resolved maps of all galaxies. PAH3.3 has clear detections in ASPECS-6 and ASPECS-14 and a weak detection with matching morphology in ASPECS-15. We do not see any robust detection in the PAH3.3 map of ASPECS-11, and similarly, in ASPECS-C20, PAH3.3 is very weak, with a marginal detection at the position of the galaxy \citep[see][]{Donnan2026b}. Note that ASPECS-11, 15, and C20 do not show robust detections in their integrated spectra \citep[see][]{Lofaro2026}.

Figure~\ref{fig:PAH-maps} presents maps of the ratio of total PAH7.7 flux to synthetic photometry performed on the MRS cubes using the F1500W filter, which at the redshift of our galaxies encompasses not only the majority (63\%) of the PAH7.7 feature but also the dust continuum underneath, part (24\%) of the PAH6.2 feature, and other fainter PAHs at $\sim\,6.7\,\mu$m and $\sim$\,7.1\,$\mu$m (see the bottom-left panel of ASPECS-6 in Figure~\ref{fig:spectra}). 
Note that the PAH7.7/F1500W can be larger than 1, since the wings of the 7.7\,$\mu$m PAH feature (especially the red wing) extend beyond the F1500W filter, which can cause the total PAH7.7 flux to be larger than the F1500W synthetic flux, even if the latter also includes continuum emission and some PAH6.2 flux. The second and third row of Figure~\ref{fig:PAH-maps} show the fraction of PAH7.7 and continuum, respectively, within the F1500W filter. That is, the PAH7.7 fraction (second row) was derived by accounting only for the 63\% of the total PAH7.7 flux that is actually covered by the F1500W filter. To derive the continuum fraction, we simply used the continuum derived from the spectral decomposition and integrated its flux within the filter. For ASPECS-15, a companion galaxy dominates the total continuum emission, so this component includes contributions from both ASPECS-15 and the companion. We note that the sum of the PAH and continuum contributions to the total flux does not add up to 100\%. This is because the F1500W filter also includes the reddest part of the wing of the 6.2\,$\mu$m PAH feature as well as the emission plateau from the other fainter PAH.

For ASPECS-6, where the PAH7.7\,$\mu$m feature is most resolved (it is the largest galaxy in the sample), the continuum and PAH maps differ more significantly. Such variations can also be seen to some extent for the other sources. In ASPECS-6, the fractional contribution of continuum emission (3rd row in Figure~\ref{fig:PAH-maps}) has a peak at the position of the galaxy's core, while the PAH emission is distributed on larger scales, suggesting that the warm dust emission is more concentrated than the PAH feature, similar to what has been found in nearby, luminous dusty galaxies \citep[e.g.,][]{DS2011}. The contribution of PAH7.7 to the F1500W filter ranges from $\sim$\,70\,\% to almost 85\%, while the continuum contributes from up to $\sim$\,5\,\% to less than 1\%. This implies that, while the total emission within the F1500W is dominated by PAH emission (mostly the 7.7\,$\mu$m feature but also other shorter-wavelength PAHs) and therefore can be used to reasonably infer scaling relations \citep[see][]{Shivaei2024}, photometric imaging alone is not sufficient to characterize the distribution and isolate the PAH7.7 emission from other dust components, thus requiring the use of IFU data.

\subsection{PAH Properties}
\label{pah-properties}

\subsubsection{PAH Mass-to-Light Ratios}

\begin{figure}[ht]
        \centering
        \includegraphics[width=\columnwidth]{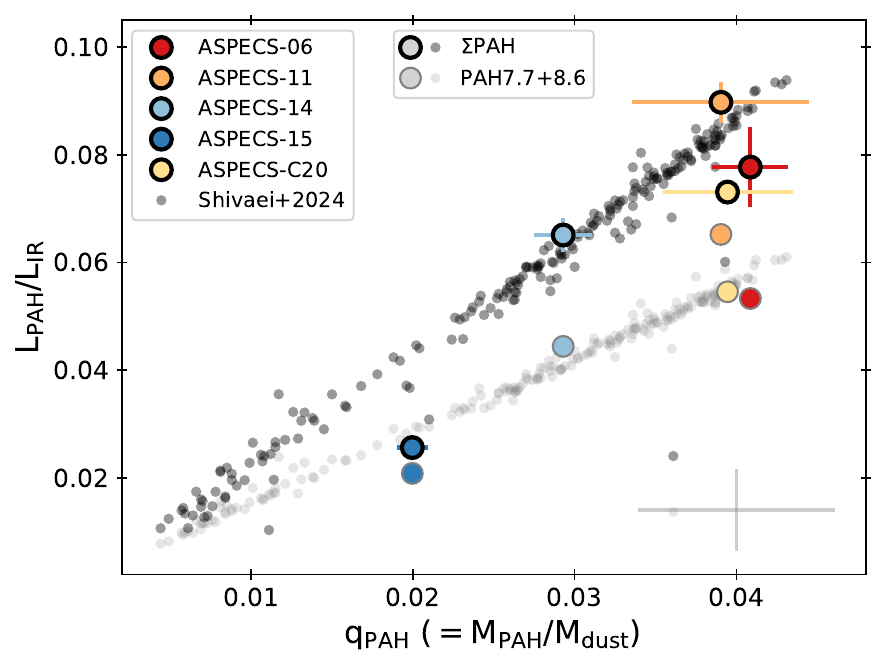} 
    \caption{Ratio of total PAH (and PAH7.7+8.6) to total IR luminosities as a function of mass fraction of PAH to total dust. PAHSPECS data are shown with different colors for each galaxy. The total PAH (combination of 3.3, 6.2, 7.7, 8.6, and 11.3\,$\mu$m bands) is shown with darker colors and the PAH7.7+8.6 with lighter. The PAH luminosities of the PAHSPECS galaxies are measured from MRS spectra and the IR luminosities are from the SED fits to UV-FIR photometry. The spectroscopic PAH luminosities are multiplied by a scaling factor accounting for the offset between the MRS spectra and the photometric data (Figure~\ref{fig:spectra}, see text and \citealt{Boogaard2026}).
    For comparison, we show data from \cite{Shivaei2024a} where both PAH and IR luminosities are derived from best-fit SEDs to UV-MIR photometry. The PAH luminosities from \cite{Shivaei2024a} are measured using the clipping methodology with corrections applied according to Table~\ref{tab:totalclip} (see text).  
    PAH mass fractions in both cases are derived from SED fitting using \cite{Draine2007} models.
    The tightness of the relation in the \cite{Shivaei2024a} data is by design, since both the L$_{\mathrm{PAH}}$ and q$_{\mathrm {PAH}}$ are derived from the same best-fit model to the photometry, while L$_{\mathrm {PAH}}$ for PAHSPECS galaxies is directly from observed spectra. The average uncertainty of this dataset is shown on the bottom-right corner.
    }
    \label{fig:PAHlum-mass}
\end{figure}

In Figure~\ref{fig:PAHlum-mass}, we compare the PAH-to-total IR luminosity ratio obtained from the MRS spectroscopy with the PAH-to-dust mass ratio, $q_{\rm PAH}$, obtained from the photometric SED fitting. The PAH luminosities are derived via \texttt{CAFE} spectral decomposition and the IR luminosity is defined as the integral over the 8--1000\,$\mu$m range. The PAH-to-dust mass ratio is inferred from the best photometric SED model, fitted to the combined MIRI, Herschel, and ALMA data. We present luminosity ratios for both the total PAH emission (3.3, 6.2, 7.7, 8.6, and 11.3\,$\mu$m combined) and the 7.7+8.6\,$\mu$m complex alone. For comparison, the gray points show the luminosity and mass ratios from \cite{Shivaei2024a}, derived for a large sample of star-forming galaxies at $z$\,$\sim$\,0.7--2.0 using only MIRI photometry.

The luminosity and mass ratios are expected to be tightly correlated in the \citet{Shivaei2024a} data, as both quantities are inferred from the \texttt{Prospector} best-fit SED models based on \cite{Draine2007}. In comparison to the PAHSPECS PAH-to-IR luminosity ratios, the SED-inferred $q_{\rm PAH}$ is overestimated for the X-ray AGN source ASPECS-15 \citep[see also][]{Boogaard2026}. ASPECS-15 also has the lowest PAH-to-continuum ratio in the sample, consistent with the presence of an AGN, making its SED fitting more challenging, even when AGN models are included in SED fitting. In addition, ASPECS-15 is the galaxy with the highest $\tau_V$, which might partially contribute to the discrepancy, as the \texttt{CAFE} fits do not account for extinction and may therefore have underestimated the PAH luminosity. 
The other PAHSPECS galaxies are in good agreement with the \citet{Shivaei2024a} trend shown in Figure~\ref{fig:PAHlum-mass}, but with a larger scatter in the spectroscopic PAH-to-IR luminosity ratio at fixed PAH-to-dust mass ratio. 
The larger scatter suggests that mass-to-light PAH and total dust/IR ratios are not as tightly correlated as in the models, probably pointing to variations in dust physical properties and/or geometries that current models cannot account for based on photometric data alone.

\subsubsection{PAH Equivalent Widths}

\begin{figure}[ht]
        \centering
        \includegraphics[width=.9\columnwidth]{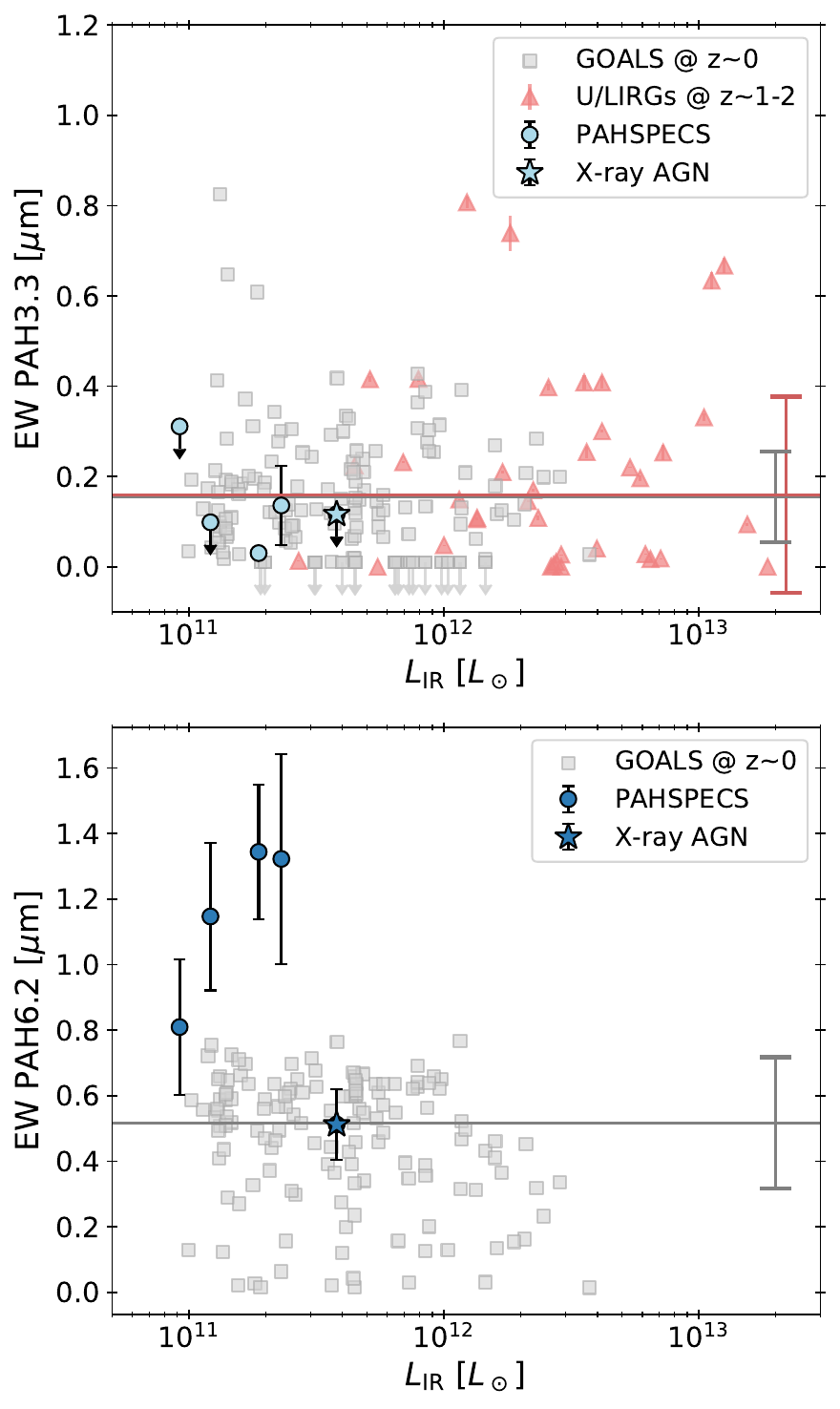} 
    \caption{Equivalent width (EW) as a function of the total IR luminosity of galaxies, for the 3.3\,$\mu$m (top) and 6.2\,$\mu$m (bottom) PAH features. PAHSPECS galaxies are shown as circles, local luminous infrared galaxies from the GOALS sample \citep{Armus2009, Stierwalt2013, Inami2018} are plotted as gray squares, and (ultra)-luminous infrared galaxies from the cosmic noon sample of starburst galaxies from \cite{McKinney2026} are displayed in light red triangles. The 3$\sigma$-clipped median and standard deviation of the GOALS detected sample (grey) and the starburst sample (red) are shown with horizontal lines and vertical error bars, respectively.
    As a population, ``normal'' PAHSPECS galaxies show PAH3.3 EWs on average smaller than nearby LIRGs (considering the data an upper limits together) and perhaps are more in common with massive, starbursting galaxies at $z$\,$\sim$\,1--2, despite the order-of-magnitude difference in their IR luminosities. On the other hand, PAH6.2 EWs of galaxies from the PAHSPECS sample are on average significantly larger than the local LIRG population.
    }
    \label{fig:PAHEW}
\end{figure}

Figure~\ref{fig:PAHEW} presents the equivalent width (EW) measured for the PAHSPECS sample (colored circles) as a function of $L_{\rm IR}$. The measurements are compared to nearby luminous infrared galaxies (LIRGs) from the GOALS sample from \cite{Armus2009} (light gray squares), and to the population of cosmic noon ($z$\,$\sim$\,0.7--2) massive, starburst galaxies from \cite{McKinney2026} (red triangles). We note that the PAH3.3 and continuum measurements for both PAHSPECS and GOALS galaxies have been derived via \texttt{CAFE} decomposition\footnote{The \texttt{CAFE} fit of GOALS galaxies is based on AKARI and Spitzer/IRS spectroscopy \citep{Inami2018}.}, while those for the cosmic noon starburst sample were obtained directly from the spectra, via the \textit{clipping} method described in the previous sections. According to Figure~9 of \cite{McKinney2026}, the ratio between PAH3.3 luminosities obtained with these two methods range from $\simeq$\,2 to up to $\sim$\,8. Assuming a factor of 2 correction for the \cite{McKinney2026} sample, the top panel of Figure~\ref{fig:PAHEW} shows that both PAHSPECS galaxies and cosmic noon starburst U/LIRGs cover the same range in PAH3.3 EWs with similar uniformity, with values in general being $\lesssim$\,0.2$\,\mu$m (starburst U/LIRGs sample median is 0.11\,$\mu$m and standard deviation of 0.06\,$\mu$m), despite the $z$\,$\sim$\,1--2 U/LIRGs having on average one order of magnitude higher $L_{\rm IR}$. Nearby luminous infrared galaxies, on the other hand, extend the higher envelope of the PAH3.3 EW distribution to values $>$\,0.2\,$\mu$m (median of 0.16\,$\mu$m with standard deviation of 0.13\,$\mu$m), suggesting that galaxies at cosmic noon, no matter whether they are normal or starbursting, show on average a reduced population of small, neutral grains when compared to the local dusty galaxy population. The implications of these results are extensively discussed in the companion papers by \cite{Lofaro2026} and \cite{Donnan2026b}.

Interestingly, when looking at the bottom panel of Figure~\ref{fig:PAHEW}, the values obtained for the PAH6.2 EW of the PAHSPECS galaxies show on average an excess, rather than a deficit, with respect to the nearby LIRG population, with 3 out of 5 PAHSPECS galaxies displaying values significantly larger ($1.5-1.8\times$) than the higher envelope of the LIRG distribution, and $2.2-2.6\times$ larger than the median of the LIRG distribution, 0.52\,$\mu$m with standard deviation of 0.20\,$\mu$m \cite[see also][]{Lofaro2026}. We note that \cite{Smith2007} report PAH6.2 EW values of up to $\sim$\,3\,$\mu$m in local, normal star-forming galaxies, but using a different decomposition tool, which complicates a direct comparison with this work. The behavior of the PAH6.2 feature is in stark contrast with what is found for the 3.3\,$\mu$m PAH, and provides a reason why the total intensity of the 6.2 and 7.7\,$\mu$m PAH features can be approximated so well by photometric SED fitting alone (see Figure~\ref{fig:PAH-maps}). That is, the PAH6.2 EW is sufficiently strong so that broad-band imaging is not significantly contaminated by dust continuum emission, in agreement with our findings in Section~\ref{sec:resolved-maps}. Examples of the spatial distribution of PAH emission and their ratios for all the galaxies in PAHSPECS can be found in \cite{Donnan2026b}.

Having in mind the limited statistics of the PAHSPECS sample, the PAH6.2 EW values found for our galaxies seem to indicate that a large fraction of normal galaxies at cosmic noon have larger PAH6.2 luminosities than local starbursts, signaling a potentially larger contribution of small dust grains from the ionized PAH population. Note that this is the opposite of what is found for the neutral population traced by the 3.3\,$\mu$m feature. The systematically low EW of the 3.3\,$\mu$m feature suggests that the neutral PAH population is skewed towards larger grains. A thorough discussion regarding this finding can be found again in \cite{Lofaro2026} and \cite{Donnan2026b}.

\section{Main Science Goals}
\label{sec:ScienceGoals}

In this section we briefly describe the main goals of the detailed studies spurred by the PAHSPECS program. Each work focuses on a particular aspect of the data, taking advantage of the spectral and spatial properties of the galaxies in the sample:

\subsection{Integrated PAH properties and ratios}

\citet{Lofaro2026} use spatially integrated JWST/MIRI MRS spectra of the five PAHSPECS galaxies to measure their PAH band luminosities and ratios, and relate them to global galaxy properties such as $L_{\rm IR}$, SFR, and specific SFR. By comparing these measurements with those of nearby LIRGs, they assess the applicability of well-established local PAH--SFR relations and PAH diagnostics to main-sequence galaxies at cosmic noon, while probing potential differences in their ISM conditions relative to the local Universe. \citet{Lofaro2026} also examine whether the AGN-hosting PAHSPECS source exhibits distinct PAH behavior relative to nearby AGN-dominated systems.

\subsection{Spatially resolved PAH and dust emission}


\cite{Donnan2026b} use the high angular resolution of MIRI-MRS along with a novel forward-modeling method to create spatially resolved maps of PAH band ratios across the ASPECS galaxies, thereby unlocking statistical studies of the evolution of resolved PAH properties at cosmic noon (see, e.g., Figure~\ref{fig:PAH-maps}). \cite{Donnan2026b} find clear differences in the radial profile of PAH properties compared to local galaxies and put forward possible scenarios to explain the newly found trends.

\subsection{PAH abundance and relation to gas and dust content}

\cite{Boogaard2026} focus on the connection between the PAHs and the global molecular gas- and dust content of the ISM of galaxies. They first use spectro-photometric modeling of the MRS and multi-wavelength data to measure the PAH-to-total dust mass fraction ($q_{\rm PAH}$) of the PAHSPECS galaxies, and compare it to values inferred from broad-band MIRI photometry alone. They also investigate how the significantly higher SFRs and ISM masses of the PAHSPECS galaxies, compared to their local counterparts at similar stellar mass, impacts the mass fraction of PAHs in connection with global ISM enrichment. Finally, they use the PAHSPECS spectroscopy to revisit the tight correlation between $L_{\rm PAH}$ and $L^\prime$(CO) that has been recently put forward for typical star-forming galaxies at cosmic noon based on MIRI multi-band photometry \citep{Shivaei2024b}.

\subsection{Spatial profiles of stars, PAHs and dust from imaging}

The simultaneous MIRI imaging of the PAHSPECS galaxies forms a deep (up to 20\,h), 15\,$\mu$m, 7 square arcmin mosaic in the HUDF that reaches unprecedented sensitivity (as deep as 5$\sigma$ depth of 26.62 mag).
Using the 15$\mu$m image as a probe of PAH rest-frame 6--9\,$\mu$m emission in tens of galaxies at cosmic noon, \cite{FA2026} study the spatially resolved $\sim 7-8\,\mu$m PAH and hot dust observations that are combined with star-formation tracers from near-IR JWST data and cold gas and dust measurements with ALMA to explore the resolved dust physics and dust-obscured star formation within typical galaxies at the peak of cosmic star formation and stellar mass assembly.

\section{Summary}
\label{sec:Summary}

We have presented PAHSPECS, an unbiased JWST/MIRI MRS survey of a complete, dust- and CO-selected sample of five normal, main-sequence galaxies at $z$\,$\sim$\,1.1 from the blind ASPECS survey in HUDF. By obtaining spatially resolved spectroscopy across the full MIRI/MRS wavelength range at 5 to $26\,\mu$m, PAHSPECS provides the first simultaneous view of the major PAH features at 3.3, 6.2, 7.7, 8.6, and 11.3\,$\mu$m in a representative sample of galaxies at cosmic noon. Our main results are summarized below.

\begin{enumerate}

\item We detect the 6.2, 7.7, and 11.3\,$\mu$m PAH features in all five galaxies, while the weaker 3.3\,$\mu$m feature is detected in two sources. We find that multi-wavelength photometric SED modeling can recover the luminosities of the strong 6.2 and 7.7\,$\mu$m features to within $\sim 20-25\%$ of those measured directly from the MIRI/MRS spectra using the \texttt{CAFE} spectral decomposition tool. In contrast, the 3.3 and 11.3\,$\mu$m features can be substantially overestimated by photometric modeling, by more than a factor of 2, and as large as a factor of 5 for the 3.3\,$\mu$m flux. This demonstrates that broadband photometry can provide reliable estimates of the strong PAH emissions, but spectroscopy remains essential for characterizing PAH band ratios and the underlying dust properties.

\item The PAH-to-dust mass fraction, $q_{\rm PAH}$, inferred from photometric SED fitting is broadly consistent with the spectroscopically measured PAH-to-IR luminosity ratio, but the correlation is less tight than predicted by models, highlighting variations in dust physical properties and/or geometries that current models do not account for based on photometric data alone.

\item The PAH equivalent widths reveal a striking difference between the neutral and ionized PAH populations at cosmic noon. The 3.3\,$\mu$m PAH feature, which predominantly traces neutral PAHs and is sensitive to the abundance of small grains, has systematically low equivalent widths compared with local luminous infrared galaxies. In contrast, the 6.2\,$\mu$m feature, which is more strongly associated with ionized PAHs, shows enhanced equivalent widths, with three of the five galaxies lying above the upper envelope of the local LIRG population. These results suggest that the PAH population in typical galaxies at cosmic noon is not simply a scaled version of that in local galaxies, but instead exhibits differential evolution in the size distributions of neutral and ionized PAHs. In particular, the weak 3.3\,$\mu$m emission is consistent with a relative deficit of small neutral PAHs, while the strong 6.2\,$\mu$m emission points to a substantial population of small ionized grains.

\item The PAH emission is spatially extended and exhibits significant variations across individual galaxies. 
For the most extended galaxy, ASPECS-6, spatially resolved spectroscopy reveals variations in the contribution of PAH7.7 emission to the MIRI F1500W band, where continuum peaks toward the galaxy core, while the PAH emission is more extended. Although deep F1500W imaging is dominated by PAH emission and can therefore serve as a useful proxy for total PAH7.7 emission, spatially resolved MIRI/MRS spectroscopy is essential to isolate the 7.7\,$\mu$m PAH feature and reveal its distribution relative to the warm dust continuum.
\end{enumerate}

Taken together, these observations and the accompanying papers \citep{Lofaro2026,Donnan2026b,Boogaard2026,FA2026} establish that PAHs at cosmic noon already exhibit a complex and evolving relationship between grain size, charge state, and the physical conditions of the ISM. The contrasting behavior of the 3.3 and 6.2\,$\mu$m features provides evidence that the neutral and ionized PAH populations respond differently to the high gas densities, radiation fields, and chemical enrichment characteristic of galaxies at this epoch. PAHSPECS therefore demonstrates the power of JWST/MIRI spectroscopy to move beyond using PAHs simply as tracers of star formation and instead use their detailed spectral and spatial properties to probe the evolution of the multiphase ISM. Larger samples and spatially resolved measurements combining PAHs with molecular gas, dust, and ionized-gas diagnostics will be crucial for determining the physical mechanisms governing PAH formation, growth, destruction, and charging across cosmic time.

\begin{acknowledgements}
Authors acknowledge fruitful discussions throughout the project and data reduction of PAHSPECS with David Law, Andreas Faisst, and Wuji Wang.
IS acknowledges funding from the Atracc\'{i}on de Talento Grant No. 2022-T1/TIC-20472 of the Comunidad de Madrid, Spain, as well as the European Research Council (ERC) under the European Union’s Horizon 2020 research and innovation programme (DistantDust, Grant agreement No. 101117541).
LAB acknowledges support from the Dutch Research Council (NWO) under grant VI.Veni.242.055 (\url{https://doi.org/10.61686/LAJVP77714}).
FRD and KS acknowledge funding support from grant JWST-GO-05279.002.
FG acknowledges support by the French National Research Agency under the contracts WIDENING (ANR-23-ESDIR-0004) and REDEEMING (ANR-24-CE31-2530), as well as by the Actions Thématiques ``Physique et Chimie du Milieu Interstellaire'' (PCMI) of CNRS/INSU, with INC and INP, and ``Cosmologie et Galaxies'' (ATCG) of CNRS/INSU, with INP and IN2P3, both programs being co-funded by CEA and CNES.
MA is supported by FONDECYT grant number 1252054, and gratefully acknowledges support from ANID Basal Project FB210003,  ANID MILENIO NCN2024\_112 and ANID + Vinculaci\'on Internacional + FOVI250261.
This work is based on observations made with the NASA/ESA/CSA James Webb Space Telescope. The data were obtained from the Mikulski Archive for Space Telescopes at the Space Telescope Science Institute, which is operated by the Association of Universities for Research in Astronomy, Inc., under NASA contract NAS 5-03127 for JWST. PAHSPECS observations are associated with program 5279. JADES DR5 includes NIRCam data from JWST programs 1176, 1180, 1181, 1210, 1264, 1283, 1286, 1287, 1895, 1963, 2079, 2198, 2514, 2516, 2674, 3215, 3577, 3990, 4540, 4762, 5398, 5997, 6434, 6511, and 6541. JADES DR5 includes MIRI data from JWST programs 1180, 1181, and 1207.
The authors acknowledge the teams of programs 1895, 1963, 2079, 2514, 3215, 3577, 3990, 6434, and 6541 for developing their observing program with a zero-exclusive-access period. 
This research is also in part based on observations made with the NASA/ESA Hubble Space Telescope obtained from the Space Telescope Science Institute, which is operated by the Association of Universities for Research in Astronomy, Inc., under NASA contract NAS 5–26555. 
This paper also makes use of the following ALMA data: program 2016.1.00324.L. ALMA is a partnership of ESO (representing its member states), NSF (USA) and NINS (Japan), together with NRC (Canada), NSTC and ASIAA (Taiwan), and KASI (Republic of Korea), in cooperation with the Republic of Chile. The Joint ALMA Observatory is operated by ESO, AUI/NRAO and NAOJ."

\end{acknowledgements}
  
\bibliographystyle{./aa}
\bibliography{./bib}{}


\appendix
\section{SEDs} \label{sec:app1}

Figures~\ref{fig:seds} and \ref{fig:seds2} show the wealth of photometric data and the best-fit SEDs of the five PAHSPECS galaxies using \texttt{Prospector} SED fitting tool. For each galaxy, we also show the NIRCam RGB cutout made from F277W, F356W, and F444W, and the deep MIRI F1500W cutout of the source from our simultanous imaging (Section~\ref{sec:miri-imaging}).

\begin{figure*}[ht]
        \centering
        \includegraphics[trim={0 0 0 5cm},clip,width=.8\textwidth]{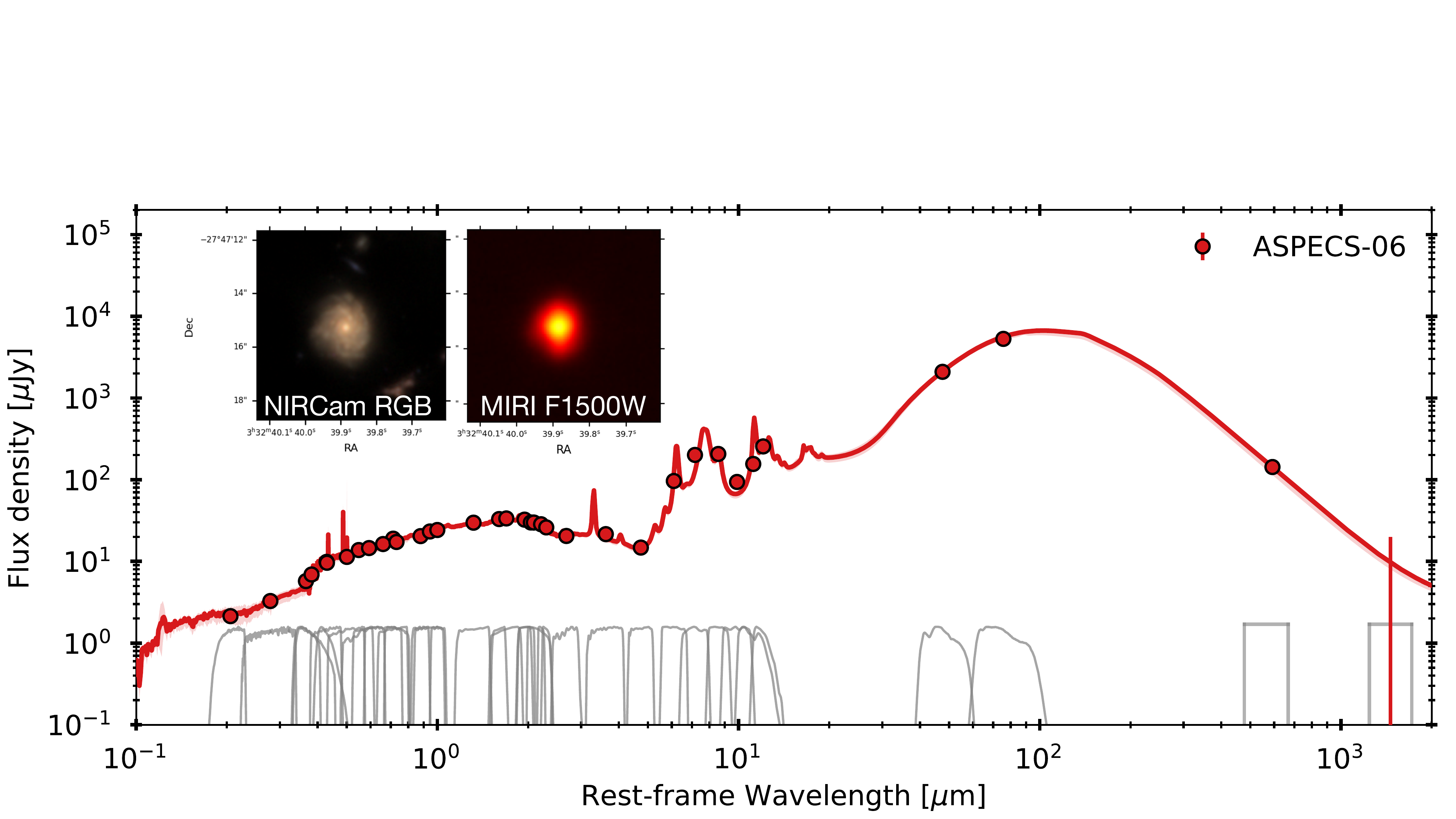} \\
        \includegraphics[trim={0 0 0 5cm},clip,width=.8\textwidth]{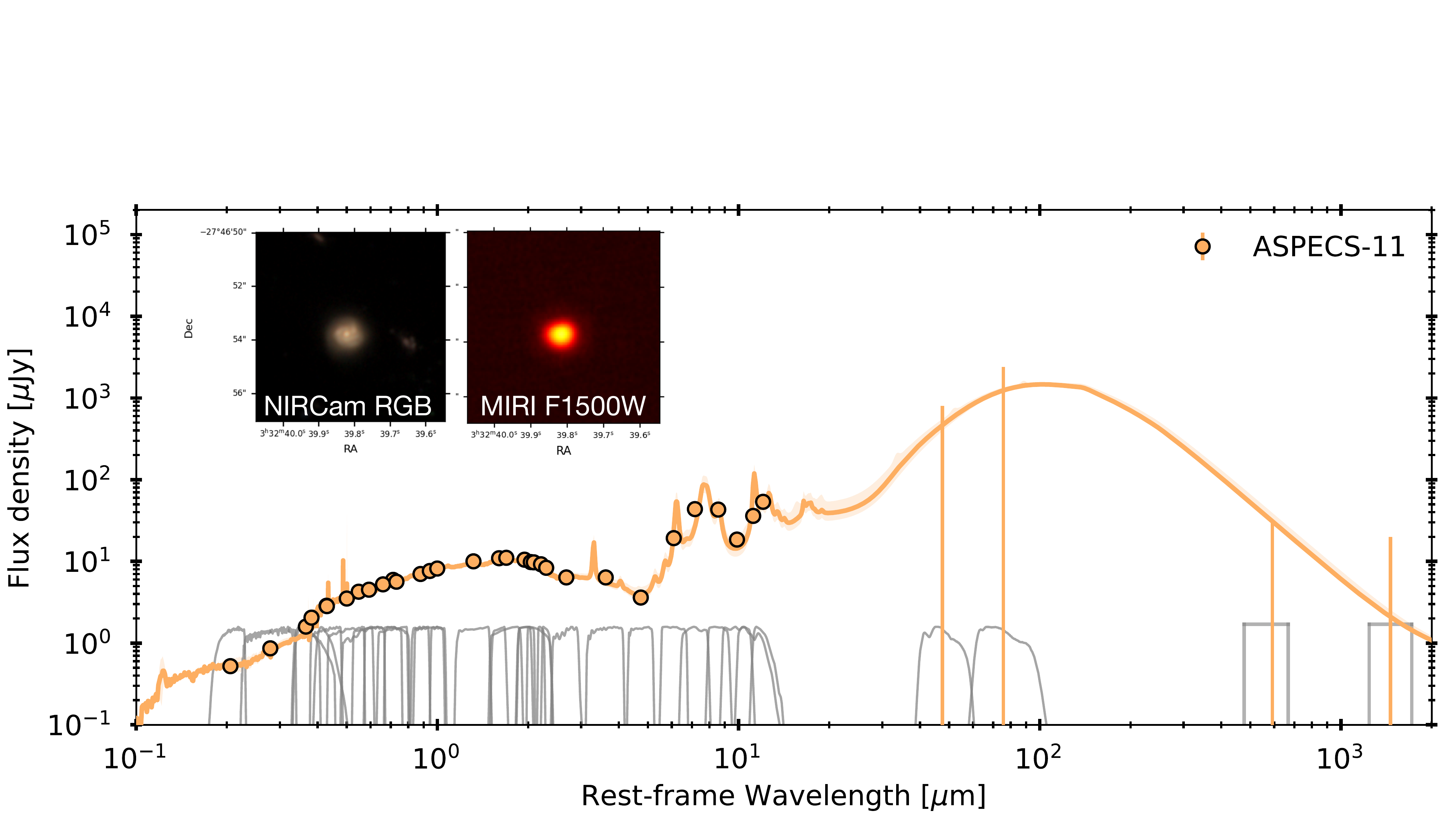} \\
        \includegraphics[trim={0 0 0 5cm},clip,width=.8\textwidth]{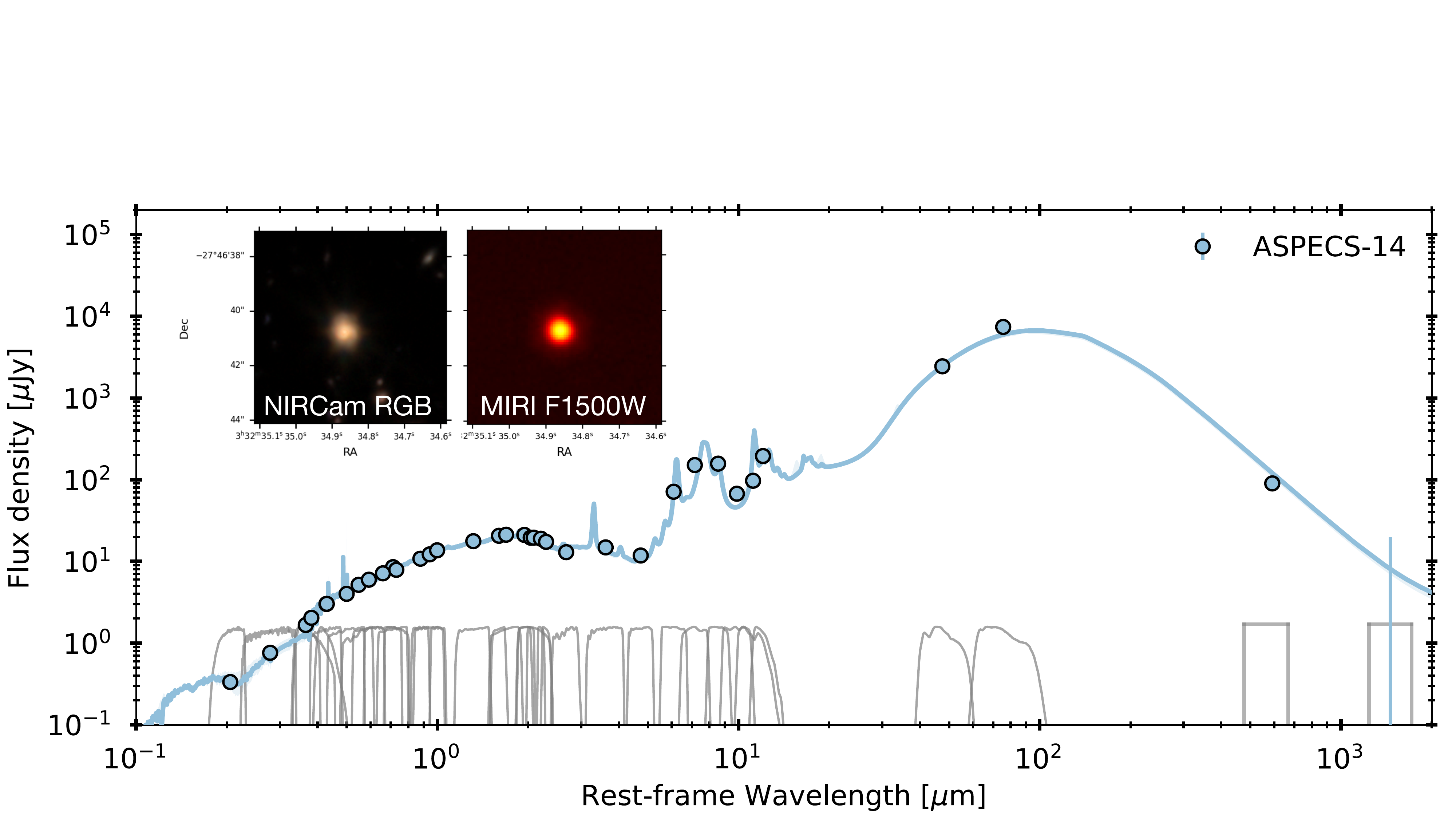} 
    \caption{Best-fit SEDs and observed photometry of PAHSPECS sample. Filter curves used in the SED fitting are shown at the bottom. 7 by 7 arcsec NIRCam and MIRI cutouts are also shown.
    NIRCam RBG cutout is from filters F277W, F356W and F444W. }
    \label{fig:seds}
\end{figure*}

\begin{figure*}[ht]
        \centering
        \includegraphics[trim={0 0 0 5cm},clip,width=.8\textwidth]{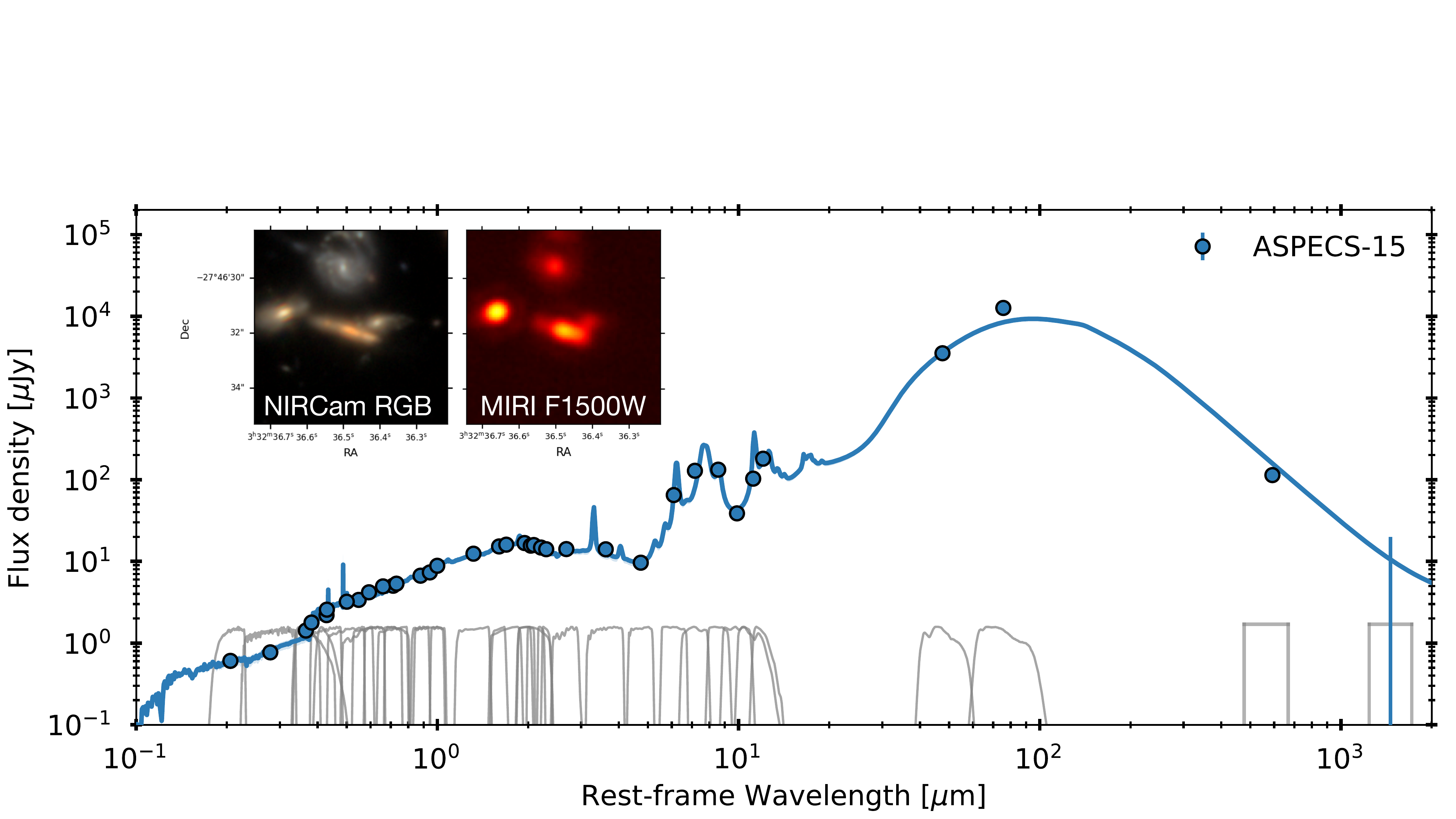} \\
        \includegraphics[trim={0 0 0 5cm},clip,width=.8\textwidth]{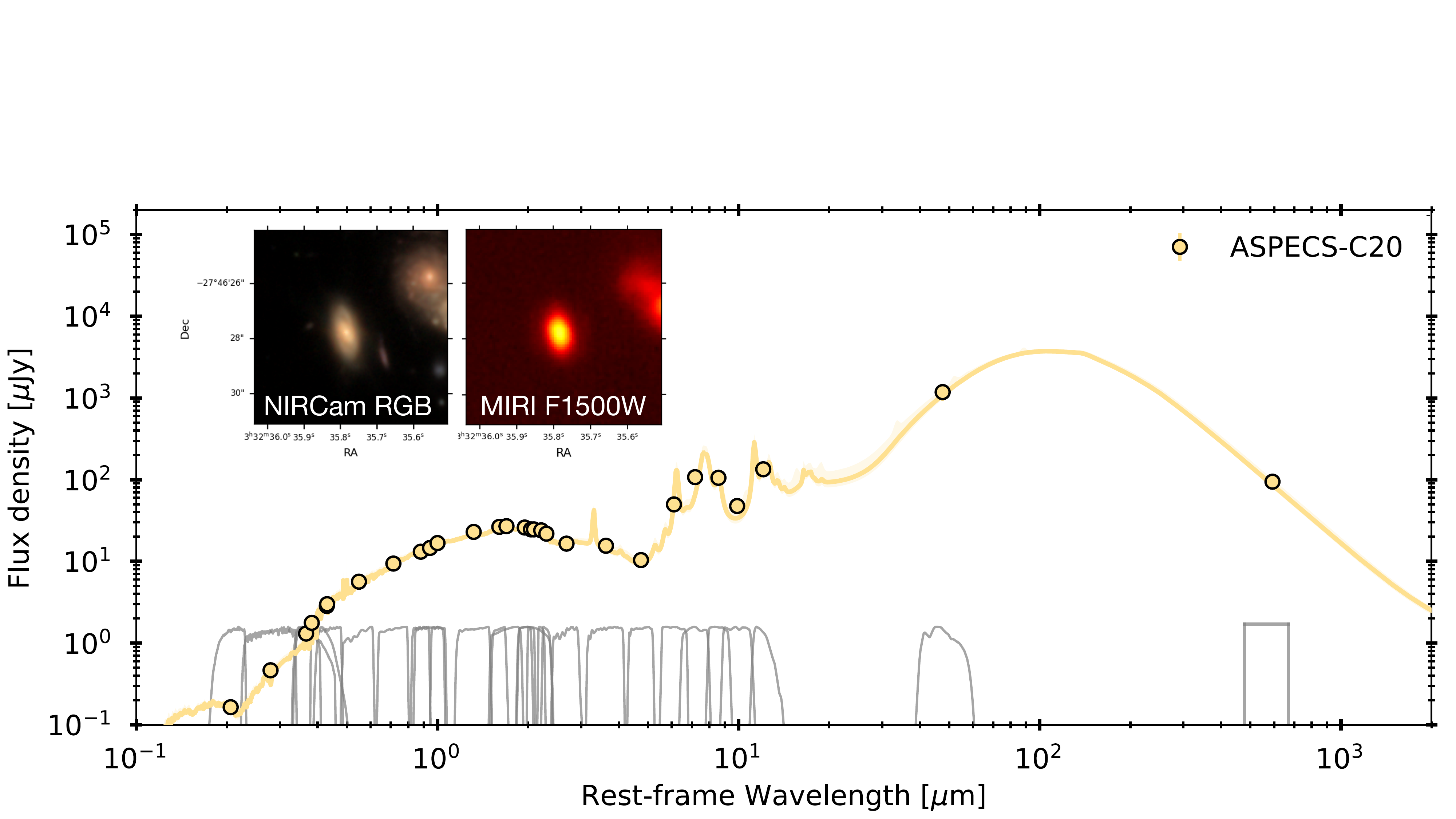} \\
    \caption{Continued from Fig.~\ref{fig:seds}}
    \label{fig:seds2}
\end{figure*}

\end{document}